\documentclass[sigconf]{acmart}
\usepackage{multirow}
\usepackage{subcaption}
\usepackage{color}
\usepackage{enumitem}
\usepackage{makecell}
\usepackage{algorithm}
\usepackage{listings}
\usepackage[table]{xcolor}
\definecolor{myblue}{RGB}{224,231,251}
\definecolor{myred}{RGB}{255,230,230}
\definecolor{textblue}{RGB}{70,93,170}
\definecolor{textred}{RGB}{255,25,25}
\setlist[itemize]{leftmargin=*}

\AtBeginDocument{%
  \providecommand\BibTeX{{%
    \normalfont B\kern-0.5em{\scshape i\kern-0.25em b}\kern-0.8em\TeX}}}

\setcopyright{acmlicensed}
\copyrightyear{2018}
\acmYear{2018}
\acmDOI{XXXXXXX.XXXXXXX}

\acmConference[Conference acronym 'XX]{Make sure to enter the correct
  conference title from your rights confirmation emai}{June 03--05,
  2018}{Woodstock, NY}
\acmBooktitle{Woodstock '18: ACM Symposium on Neural Gaze Detection,
 June 03--05, 2018, Woodstock, NY} 
\acmISBN{978-1-4503-XXXX-X/18/06}

\begin{document}

% \title{Not Everything Needs to be Forgotten:\\Recommendation Unlearning via Guided Filtering}
% \title{Forgetting the Outdated: Efficient Recommendation Unlearning via Guided Filtering}
% \title{Retaining Beneficial Knowledge:\\Recommendation Unlearning via Guided Filtering}
% \title{Retaining Makes Better:\\Recommendation Unlearning via Guided Filtering}
% \title{Preserving Makes Better:\\Recommendation Unlearning via Guided Filtering}
% \title{Filtering User Preferences from Recommendation Unlearning}
% \title{An Efficient Guided Filtering Framework to Erase Outdated Preferences for Recommendation Unlearning}
% \title{Pay Attention to Data Split: A Reproducibility Study of Sequential Recommendation Models with Sub-Sequence Splitting}
\title{Pay Attention to Sequence Split: Uncovering the Impacts of Sub-Sequence Splitting on Sequential Recommendation Models}

\author{Yizhou Dang}
\affiliation{%
  \institution{Software College, Northeastern University}
  \city{Shenyang}
  \country{China}
}
\email{dangyz@mails.neu.edu.cn}

\author{Yifan Wu}
\affiliation{%
  \institution{Software College, Northeastern University}
  \city{Shenyang}
  \country{China}
}
\email{wuyf4@mails.neu.edu.cn}

\author{Minhan Huang}
\affiliation{%
  \institution{Software College, Northeastern University}
  \city{Shenyang}
  \country{China}
}
\email{huangmh2@mails.neu.edu.cn}

\author{Chuang Zhao}
\affiliation{%
  \institution{Tianjin University}
  \city{Tianjin}
  \country{China}
}
\email{zhaochuang@tju.edu.cn}

\author{Lianbo Ma}
\affiliation{%
  \institution{Software College, Northeastern University}
  \city{Shenyang}
  \country{China}
}
\email{malb@swc.neu.edu.cn}

\author{Guibing Guo}
\authornote{Corresponding author.}
\affiliation{%
  \institution{Software College, Northeastern University}
  \city{Shenyang}
  \country{China}
}
\email{guogb@swc.neu.edu.cn}

\author{Xingwei Wang}
\authornotemark[1]
\affiliation{%
  \institution{School of Computer Science and Engineering, Northeastern University}
  \city{Shenyang}
  \country{China}
}
\email{wangxw@mail.neu.edu.cn}

\author{Zhu Sun}
\affiliation{%
  \institution{Information Systems Technology and Design, Singapore University of Technology and Design}
  \city{Singapore}
  \country{Singapore}
}
\email{sunzhuntu@gmail.com}

\renewcommand{\shortauthors}{Yizhou Dang, et al.}

\begin{abstract}
Sub-sequence splitting (SSS) has been demonstrated as an effective approach to mitigate data sparsity in sequential recommendation (SR) by splitting a raw user interaction sequence into multiple sub-sequences. Previous studies have demonstrated its ability to enhance the performance of SR models significantly. However, in this work, we discover that \textbf{(i). SSS may interfere with the evaluation of the model's actual performance.} We observed that many recent state-of-the-art SR models employ SSS during the data reading stage (not mentioned in the papers). When we removed this operation, performance significantly declined, even falling below that of earlier classical SR models. The varying improvements achieved by SSS and different splitting methods across different models prompt us to analyze further when SSS proves effective. We find that \textbf{(ii). SSS demonstrates strong capabilities only when specific splitting methods, target strategies, and loss functions are used together.} Inappropriate combinations may even harm performance. Furthermore, we analyze why sub-sequence splitting yields such remarkable performance gains and find that \textbf{(iii). it evens out the distribution of training data while increasing the likelihood that different items are targeted.} Finally, we provide suggestions for overcoming SSS interference, along with a discussion on data augmentation methods and future directions. We hope this work will prompt the broader community to re-examine the impact of data splitting on SR and promote fairer, more rigorous model evaluation. All analysis code and data will be made available upon acceptance. We provide a simple, anonymous implementation at \url{https://github.com/KingGugu/SSS4SR}.
\end{abstract}







\maketitle

\section{Introduction}

Sequential recommendation learns user preferences from their historical interaction sequences and generates personalized suggestions \cite{kang2018self,lan2023spatio,hidasi2015session,dang2023ticoserec}. Over the past several years, numerous SR models based on different architectures have been proposed, achieving remarkable progress \cite{zhou2022filter,dang2023ticoserec,yue2024linear,di2025personalized}. However, most users tend to interact with only a small number of items on the platform, while most items receive very few interactions \cite{yin2020overcoming,dang2024augmenting,di2025federated}. Consequently, the widespread data sparsity limits the models' performance. To tackle this, researchers proposed many effective data augmentation (DA) methods to generate additional training data \cite{liu2023diffusion,liu2021augmenting,dang2024data,liu2024llm}.

\begin{figure}[!t]
  \centering
  \includegraphics[scale=0.535]{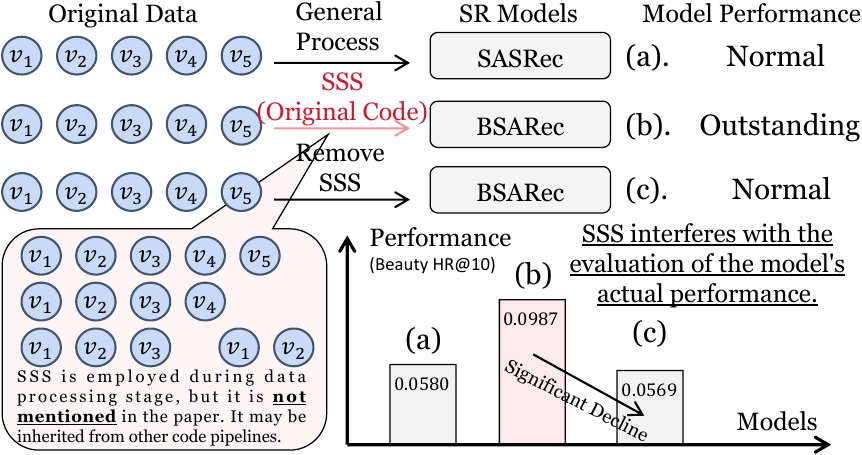}
  \vspace{-1em}
  \caption{An example of SSS interferes with model evaluation.}
  \vspace{-1em}
  \label{fig:example}
\end{figure}

Data augmentation alleviates data sparsity by perturbing original data or directly generating new samples \cite{sun2021does,sun2025llm4rsr}. Research in this area can be broadly categorized into two types: the first involves heuristic operations based on rules and human prior knowledge, such as sub-sequence splitting (also known as sliding windows) \cite{tang2018personalized,lee2025sequential}, which split one original sequence into many sub-sequences (e.g., $[1,2,3,4] \rightarrow \{[1,2],[1,2,3],[1,2,3,4]\}$), dropout \cite{tan2016improved}, cropping \cite{xie2022contrastive}, masking \cite{xie2022contrastive}, and insert \cite{liu2021contrastive}. The second category comprises methods that leverage learnable augmentation modules, such as extending short sequences with bidirectional transformers \cite{liu2021augmenting,jiang2025improving}, replacing items by counterfactual thinking \cite{wang2021counterfactual}, and directly generating new sequences with diffusion models \cite{liu2023diffusion,wu2023diff4rec}. Although DA methods are intended to better facilitate model learning of user preferences and improve recommendation accuracy, we find that sub-sequence splitting, one of the earliest proposed and proven-effective DA methods \cite{zhou2024contrastive}, may not have been applied appropriately in many recent SR models. This problem has led to widespread issues, including misjudged performance and unfair evaluations.

\vspace{0.3em}

\noindent \underline{(i). SSS Interferes with Model Evaluation. (Section \ref{sec:effects})} Through an empirical study conducted on many SR models recently published in top-tier venues, we reveal that \textbf{Sub-sequence splitting may interfere with our evaluation of the SR models' actual performance.} As shown in Figure \ref{fig:example}, we find that many recently proposed representative SR models employed SSS during data reading without acknowledging it in their papers. When we removed their SSS operations, the models' performance significantly deteriorated. Among the ten models we investigated, seven experienced performance drops exceeding 40\%. Some models even underperformed the classic model SASRec without any data augmentation \cite{kang2018self}. This result indicates that SSS interfered with researchers' accurate and rigorous evaluation of model performance. The additional performance gains it introduced led to an unfair comparison. Through investigation, we discover that this phenomenon likely stems from an earlier representative SR model that incorporated SSS into its code pipeline without mentioning it \cite{zhou2022filter}. Subsequently, many researchers implemented their own models on this pipeline. Only a handful of researchers detected the SSS and removed it.

\vspace{0.3em}

\noindent \underline{(ii). When is SSS Effective? (Section \ref{sec:when})} Furthermore, we observed that SSS yields varying degrees of improvement across different models. Given differences in loss functions (cross-entropy and binary cross-entropy), target strategies (single-target and multi-target), and splitting methods (prefix splitting, suffix splitting, and sliding window) across these models, we further explored the conditions under which SSS achieves its maximum effectiveness. Through experiments on the four most representative and influential SR models with real-world datasets, we find that \textbf{SSS demonstrates strong capabilities only when specific splitting methods, target strategies, and loss functions are used together.} Performance differences under the same model with different settings are significant. We also find that when employing appropriate settings, the early classical SR model, SASRec, can achieve performance comparable to recently proposed SR models. This observation further validates the findings in (i), namely that SSS interferes with our evaluation of the actual performance of the SR model.

\vspace{0.3em}

\noindent \underline{(iii). Why is SSS Effective? (Section \ref{sec:why})} Inspired by previous research on data augmentation \cite{dang2024repeated,lee2025sequential}, we analyze why sub-sequence splitting yields such remarkable performance gains and find that \textbf{it evens out the distribution of training data while increasing the likelihood that different items are targeted.} After adopting the sub-sequence splitting approach, both the absolute frequency and probability of low-frequency and less popular items being included in training increased. This enables the model to learn more accurate and comprehensive item representations.

\vspace{0.3em}

\noindent \underline{Discussions. (Section \ref{sec:discussion})} Based on the above analysis, we provide practical suggestions for overcoming SSS interference. When implementing and evaluating SR models, we should ensure that data splitting is applied consistently across all models. We can also choose to comprehensively evaluate the model across different settings. Additionally, we provide discussions for developing more effective DA methods, the impacts of this work beyond SR, and future directions. All analysis code and data will be made available upon acceptance. We provide a simple, anonymous implementation at \url{https://anonymous.4open.science/r/SSS_Review-45C3}.

\section{Preliminaries}

\subsection{Problem Formulation}

Suppose we have user set $\mathcal{U}$ and item set $\mathcal{V}$. Each user $u \in \mathcal{U}$ is associated with a sequence of interacted items in chronological order $s_u=[v_1, \ldots, v_j, \ldots, v_{\left|s_u\right|}]$, where $v_j \in \mathcal{V}$ indicate the item that user $u$ has interacted with at time step $j$ and $\left|s_u\right|$ is the sequence length. Given the sequences of interacted items $s_u$, sequential recommendation aims to accurately predict the most possible item $v^{*}$ that user $u$ will interact with at time step $\left|s_u\right|+1$, formulated as:
\begin{equation}
    \underset{v^{*} \in \mathcal{V}}{\arg \max} \;\; P\left(v_{\left|s_u\right|+1}=v^{*} \mid s_u \right).
\end{equation}
This equation can be interpreted as calculating the probability of all candidate items and selecting the highest one for recommendation.

\subsection{Sub-Sequence Splitting}
\label{sec:sss}
Existing studies typically employ the leave-one-out strategy, wherein the last item of each sequence serves as the test data, the items preceding it as validation data, and the remaining data as training data \cite{kang2018self,zhou2022filter,lee2025sequential}. We refer to the training data obtained by leave-one-out strategy as the original sequence $s_u$, and the sub-sequence splitting is also performed on this training data. Given a raw sequence, the following three common splitting methods exist:

\vspace{0.3em}
 
\noindent \textbf{Prefix Splitting.} It starts from the first interaction $v_1$ in the original sequence $s_u$ and divides the sequence into all possible lengths of contiguous sub-sequences. This process can be formulate as:
\begin{equation}
    \left\{s_u^{{1:2}},s_u^{1:3},\ldots,s_u^{1:|s_u|} \right\} = \operatorname{Prefix-SSS}\left(s_u\right),
\end{equation}
where $s_u^{i:j} = [v_i, v_{i+1}, \ldots, v_j]$. After the prefix SSS, we can obtain $|s_u|$ sub-sequences. For example, given an original sequence $[1,2,3,4]$, the sub-sequences after refix splitting are $\{[1,2],[1,2,3],[1,2,3,4]\}$.

\vspace{0.3em}

\noindent \textbf{Suffix Splitting.} It starts from the last interaction $v_{|s_u|}$ in the original sequence $s_u$ and divides the sequence into all possible lengths of contiguous sub-sequences. This process can be formulate as:
\begin{equation}
    \left\{s_u^{{|s_u|-1:|s_u|}},s_u^{|s_u|-2:|s_u|},\dots,s_u^{1:|s_u|} \right\} = \operatorname{Suffix-SSS}\left(s_u\right).
\end{equation}
We can also obtain $|s_u|$ sub-sequences after the suffix splitting. For example, given original sequence $[1,2,3,4]$, the sub-sequences after suffix splitting are $\{[3,4],[2,3,4],[1,2,3,4]\}$.

\vspace{0.3em}

\noindent \textbf{Sliding Window.} Given a window length $T$ and $T < \left|s_u\right| $, it divides the original sequence into multiple sub-sequences by sliding a window from one end to the other \cite{tang2018personalized}:
\begin{equation}
    \left\{s_u^{{1:T}},s_u^{2:T+1},\ldots,s_u^{|s_u|-T+1:|s_u|} \right\} = \operatorname{Sliding-SSS}\left(s_u,T\right).
\end{equation}
For example, given original sequence $[1,2,3,4,5]$ with $T=3$, the sub-sequences after suffix splitting are $\{[1,2,3],[2,3,4],[3,4,5]\}$.

\subsection{Target Strategy}
\label{sec:target}

There are two commonly used training target strategies in SR.

\vspace{0.3em}

\noindent \textbf{Single-Target Strategy.} It treats the last item in each sequence as the target for prediction during training. Therefore, it can also be referred to as the last-target strategy. For each user $u \in \mathcal{U}$, the input is all but the last item, and the target is the last item \cite{lee2025sequential}:
\begin{equation}
    x_{u}=\left[v_1, \ldots, v_{\left|s_u\right|-1}\right], \quad y_u=v_{\left|s_u\right|} .
\end{equation}
The resulting training set for user $u$ is $\mathcal{D}^{\text {train }}_{u}=\left\{\left(x_{u}, y_u\right)\right\}$.

\vspace{0.3em}

\noindent \textbf{Multi-Target Strategy.} It predicts multiple targets within each sequence. For each position $k \in\left\{2, \ldots,\left|s_{u}\right|\right\}$, the first $k-1$ items are used as the input and the $k^{\text {th}}$ item as the target \cite{lee2025sequential}:
\begin{equation}
    x^k_{u}=\left[v_1, \ldots, v_{k-1}\right], \quad y^k_{u}=v_k.
\end{equation}
This produces $\left|s_{u}\right|-1$ input-target pairs for each user sequence:
\begin{equation}
    \mathcal{D}^{\text {train }}_{u}=\left\{\left(x^k_{u}, y^k_{u}\right)\left|2 \leq k \leq\left|s_{u}\right|\right\} .\right.
\end{equation}
When $k$ equals only $\left|s_{u}\right|$, the multi-target strategy degenerates into a single-target strategy. The multi-target strategy provides more supervision signals from earlier positions in the sequence. It should be noted that prefix splitting and the multi-target strategy are orthogonal. The former involves splitting the original sequence into multiple sub-sequences, while the latter refers to a training strategy applied to a single sequence.

\section{Effects of Sub-Sequence Splitting}\label{sec:effects}

In this section, we present empirical research demonstrating how SSS influences evaluation of SR models' actual performance.

\begin{table}[!t]
  \centering
  \caption{Statistical data for the papers we investigated.}
  \vspace{-1em}
  \scalebox{0.825}{
    \begin{tabular}{c|cc}
    \toprule
    \multicolumn{1}{c|}{SSS is not used} & \multicolumn{2}{c}{SSS is used} \\
\cmidrule{2-3}    \multicolumn{1}{r|}{} & Mentioned in the Paper & \multicolumn{1}{c}{Not Mentioned in the Paper} \\
    \midrule
    \multicolumn{1}{c|}{6 papers} & 1 paper & \multicolumn{1}{c}{10 papers} \\
    \midrule
    \cite{fan2022sequential} \cite{fu2025time} \cite{neupane2024evidential} & \multicolumn{1}{c}{\multirow{2}[2]{*}{\cite{xia2025oracle}}} & \cite{zhou2022filter} \cite{li2023diffurec} \cite{liu2023distribution} \cite{shin2024attentive} \cite{baek2025muffin} \\
    \cite{xu2025multi} \cite{zhang2023adaptive} \cite{liu2024probabilistic} &       & \cite{liu2025csrec} \cite{heo2025waverec} \cite{shin2025tv} \cite{he2025exploiting} \cite{xu2025wavelet} \\
    \bottomrule
    \end{tabular}}%
  \label{tab:papers}%
  \vspace{-1em}
\end{table}%

\subsection{Identification of SR Models}
To avoid cherry-picking individual models in our study, we systematically scan the proceedings of scientific conferences and journal publications using an approach similar to the previous works \cite{ferrari2019we,shehzad2025worrying}. Specifically, we include papers in our analysis that appeared between 2022 and 2026\footnote{To enhance the timeliness of our research, we include conferences scheduled for 2026 but have already made their acceptance lists available.} in the following eight conference series: KDD, SIGIR, WWW, NeurIPS, AAAI, WSDM, CIKM, and RecSys\footnote{All of the conferences are either considered A* or A in the Australian Core Ranking.}. The journals include TOIS and TKDE. In the conferences and journals, we screened papers according to the following criteria:
\begin{itemize}
    \item The paper focuses on Classic sequential recommendation, which is the primary direction of our investigation in this work. Subdomains such as cross-domain SR \cite{zhao2023cross,zhao2023sequential}, multi-modal SR \cite{bian2023multi}, and multi-behavior SR \cite{xia2022multi}, etc., will not be considered.
    \item The primary contribution of the paper is the proposal of a foundational architecture or backbone for SR. Papers that primarily introduce auxiliary tasks (such as contrastive learning \cite{liu2021contrastive,xie2022contrastive}, reconstruction tasks \cite{ye2023graph,yin2024dataset}, alignment tasks \cite{li2024calrec}) or propose other data augmentation methods will be excluded \cite{liu2023diffusion,dang2025augmenting}\footnote{Some studies employed the open-source framework RecBole \cite{zhao2021recbole} for experimentation. However, according to RecBole's official documentation, it enables data augmentation by default. To ensure more rigorous and fair analysis, these papers are also excluded.}. This condition is designed to eliminate contributions from other auxiliary modules and tasks to the model's final performance, thereby rigorously assessing the impact of SSS on performance.
    \item Providing publicly available codes for reproduction. If these codes are not already publicly available, we contacted all paper authors and waited 30 days for a response. We select reproducible code based on the conditions established in previous work \cite{ferrari2019we}, i.e., either a working version of the source code is available or the code only needs to be modified in minimal ways to work correctly.
\end{itemize}

Applying the above criteria yielded 17 papers. It should be noted that we focus on the impact of SSS on evaluating the actual performance of the model. Therefore, reproducibility studies on how many papers provide or do not provide code, or whether they provide complete code, are not within the scope of our discussion. We conduct a thorough review and examination of the complete texts, appendices, and publicly available code of these 17 papers to determine whether they indicate the use of SSS in their papers or employ it in their code. The results are presented in Table \ref{tab:papers}. We also give a real code example of using SSS during data reading phase in Algorithm \ref{alg:code}. We observe that 6 papers make no mention of SSS and do not utilize it in their code. Among the 11 papers employing SSS, only 1 explicitly incorporates SSS as part of its methodology and analyzes it in an ablation study. The remaining ten papers use SSS in their code but fail to mention it in either the main text or the appendices. After carefully re-examining the code in these ten papers, we found their overall pipelines to be remarkably similar. Therefore, we speculate that the later works were likely based on the earlier pipeline from \cite{zhou2022filter} or on earlier pipeline improvements. Next, we will demonstrate how SSS influences researchers' evaluation of the actual performance of the methods proposed in these ten papers published in top-tier conferences \& journals.

\begin{algorithm}[!t]
  \caption{SSS code from the official implementation of \cite{zhou2022filter}}
  \label{alg:code}
  \definecolor{codeblue}{rgb}{0.25,0.5,0.5}
  \definecolor{codekw}{rgb}{0.85, 0.18, 0.50}
  \lstset{
    backgroundcolor=\color{white},
    basicstyle=\fontsize{7.5pt}{7.5pt}\ttfamily\selectfont,
    columns=fullflexible,
    breaklines=true,
    captionpos=b,
    commentstyle=\fontsize{7.5pt}{7.5pt}\color{codeblue},
    keywordstyle=\fontsize{7.5pt}{7.5pt}\color{codekw},
  }
  \begin{lstlisting}[language=python, mathescape]
  # user_seq: A list containing all original sequences.
  # max_len: Maximum Sequence Length N.

  for seq in user_seq:
    input_ids = seq[-(self.max_len + 2):-2]
    for i in range(len(input_ids)):
        self.user_seq.append(input_ids[:i + 1])

  # Then the new user_seq will be used as training data.
  \end{lstlisting}
\end{algorithm}

\begin{table}[!t]
  \centering
 \caption{Statistics of datasets. The `AL' denotes average length.}
 \vspace{-1em}
 \scalebox{0.9}{
    \begin{tabular}{c|ccccc}
    \toprule
    \textbf{Dataset} & \textbf{\# Users} & \textbf{\# Items} & \textbf{\# Interactions} & \textbf{\# AL} & \textbf{Sparsity} \\
    \midrule
    Beauty & 22,363 & 12,101 & 198,502 & 8.9   & 99.93\% \\
    Sports & 35,598 & 18,357 & 296,337 & 8.3   & 99.95\% \\
    Douyin & 20,398 & 8,299 & 139,834 & 6.9   & 99.92\% \\
    LastFM & 1,090 & 3,646 & 52,551 & 48.2  & 98.68\% \\
    \bottomrule
    \end{tabular}}%
  \label{tab:datasets}%
  \vspace{-1em}
\end{table}%

\begin{table*}[!t]
  \centering
  \caption{Performance comparison of baseline SR models and our investigated models. The `w/o SSS' indicates removing SSS from the official implementations of the corresponding models. The `Ave. Imp.' indicate the average improvements after removing SSS across four datasets. Results where performance falls below SASRec are highlighted in \textcolor{textblue}{blue}.}
    \vspace{-1em}
    \renewcommand\arraystretch{0.95}
    \setlength{\tabcolsep}{1.1mm}{
      \scalebox{0.8}{
    \begin{tabular}{cc|cccc|cccc|cccc|cccc|c}
    \toprule
    \multirow{2}[2]{*}{Model} & \multirow{2}[2]{*}{Venue} & \multicolumn{4}{c|}{Beauty}   & \multicolumn{4}{c|}{Sports}   & \multicolumn{4}{c|}{Douyin}   & \multicolumn{4}{c|}{LastFM}   & \multirow{2}[2]{*}{Ave. Imp.} \\
          &       & H@10  & N@10  & H@20  & N@20  & H@10  & N@10  & H@20  & N@20  & H@10  & N@10  & H@20  & N@20  & H@10  & N@10  & H@20  & N@20  &  \\
    \midrule
    \multicolumn{19}{c}{SR Models without SSS} \\
    \midrule
    GRU4Rec \cite{hidasi2015session} & ICLR 2016 & 0.0423  & 0.0209  & 0.0704  & 0.0279  & 0.0180  & 0.0085  & 0.0311  & 0.0118  & 0.1060  & 0.0572  & 0.1480  & 0.0678  & 0.0385  & 0.0214  & 0.0569  & 0.0260  & \multicolumn{1}{c}{—} \\
    SASRec \cite{kang2018self} & ICDM 2018 & 0.0580  & 0.0301  & 0.0910  & 0.0384  & 0.0323  & 0.0179  & 0.0490  & 0.0222  & 0.1243  & 0.0739  & 0.1677  & 0.0848  & 0.0661  & 0.0347  & 0.0982  & 0.0428  & \multicolumn{1}{c}{—} \\
    NextItNet \cite{yuan2019simple} & WSDM 2019 & 0.0324  & 0.0159  & 0.0526  & 0.0211  & 0.0169  & 0.0081  & 0.0305  & 0.0115  & 0.0696 & 0.0372 & 0.1058 & 0.0464 & 0.0544 & 0.0293 & 0.0969 & 0.0399 & \multicolumn{1}{c}{—} \\
    LRURec \cite{yue2024linear} & WSDM 2024 & 0.0602 & 0.0310  & 0.0948 & 0.0396 & 0.0330  & 0.0185 & 0.0519 & 0.0232 & 0.1239 & 0.0742 & 0.1722 & 0.0856  & 0.0650  & 0.0339 & 0.0958 & 0.0412 & \multicolumn{1}{c}{—} \\
    \midrule
    \multicolumn{19}{c}{SR Models with Contrastive Learning (The asterisk * Represents the use of SSS as part of the method for constructing contrastive views)} \\
    \midrule
    CL4SRec \cite{xie2022contrastive} & ICDE 2022 & 0.0721 & 0.0399 & 0.1048 & 0.0481 & 0.0378 & 0.0206 & 0.0554 & 0.0248 & 0.1266 & 0.0723 & 0.1707 & 0.0851 & 0.0628 & 0.0358 & 0.0976 & 0.0421  & \multicolumn{1}{c}{—} \\
    CoSeRec \cite{liu2021contrastive} & arXiv 2021 & 0.0769 & 0.0420  & 0.1102  & 0.0503 & 0.0409  & 0.0225 & 0.0613 & 0.0274 & 0.1301 & 0.0742 & 0.1785 & 0.0869  & 0.0650  & 0.0366 & 0.1001 & 0.0434  & \multicolumn{1}{c}{—} \\
    DuoRec* \cite{qiu2022contrastive} & WSDM 2022 & 0.0844 & 0.0439 & 0.1208 & 0.0531 & 0.0456 & 0.0240  & 0.0669 & 0.0293 & 0.1372 & 0.0779 & 0.1894 & 0.0912 & 0.0587 & 0.0348 & 0.0881 & 0.0422 & \multicolumn{1}{c}{—} \\
    ICSRec* \cite{qin2024intent} & WSDM 2024 & 0.0930  & 0.0562 & 0.1271 & 0.0648 & 0.0548 & 0.0318 & 0.0770  & 0.0374 & 0.1429 & 0.0823 & 0.1976 & 0.0943 & 0.0679 & 0.0415 & 0.1018 & 0.0501 & \multicolumn{1}{c}{—} \\
    \midrule
    \multicolumn{19}{c}{SR Models with SSS (Using SSS in the official code implementation but not mentioned in the paper)} \\
    \midrule
    FMLPRec \cite{zhou2022filter} & \multicolumn{1}{c|}{\multirow{2}[2]{*}{WWW 2022}} & 0.0620  & 0.0331  & \cellcolor{myblue}{0.0908} & 0.0403  & \cellcolor{myblue}{0.0309} & \cellcolor{myblue}{0.0161} & \cellcolor{myblue}{0.0474} & \cellcolor{myblue}{0.0203} & 0.1245  & \cellcolor{myblue}{0.0706} & 0.1714  & \cellcolor{myblue}{0.0825} & \cellcolor{myblue}{0.0633} & 0.0375  & 0.0982  & 0.0461  & \multicolumn{1}{c}{—} \\
    w/o SSS &       & 0.0607  & 0.0331  & 0.0926  & 0.0411  & 0.0326  & \cellcolor{myblue}{0.0175} & 0.0507  & \cellcolor{myblue}{0.0221} & 0.1316  & 0.0758  & 0.1796  & 0.0879  & \cellcolor{myblue}{0.0642} & 0.0366  & 0.0982  & 0.0452  & 3.34\% \\
    \midrule
    DLFSRec \cite{liu2023distribution} & \multicolumn{1}{c|}{\multirow{2}[2]{*}{RecSys 2023}} & 0.0658 & 0.0347 & 0.1019 & 0.0439 & 0.0394 & 0.0216 & 0.0570  & 0.0259 & 0.1523 & 0.0873 & 0.1968 & 0.0981 & 0.0721 & 0.0427 & 0.1060  & 0.0503 & \multicolumn{1}{c}{—} \\
    w/o SSS &       & 0.0593 & 0.0306 & \cellcolor{myblue}{0.0899} & 0.0396 & \cellcolor{myblue}{0.0311} & \cellcolor{myblue}{0.0173} & \cellcolor{myblue}{0.0479} & \cellcolor{myblue}{0.0218} & \cellcolor{myblue}{0.1240} & 0.0767 & 0.1779 & 0.0873 & \cellcolor{myblue}{0.0625} & 0.0371 & \cellcolor{myblue}{0.0961} & 0.0447 & -13.39\% \\
    \midrule
    DiffuRec \cite{li2023diffurec} & \multicolumn{1}{c|}{\multirow{2}[2]{*}{TOIS 2023}} & 0.0736 & 0.0434 & 0.1041 & 0.0514 & 0.0416 & 0.0227 & 0.0598 & 0.0275 & 0.1541 & 0.0909 & 0.2062 & 0.1045 & 0.0729 & 0.0418 & 0.1089 & 0.0487 & \multicolumn{1}{c}{—} \\
    w/o SSS &       & \cellcolor{myblue}{0.0374} & \cellcolor{myblue}{0.0241} & \cellcolor{myblue}{0.0504} & \cellcolor{myblue}{0.0274} & \cellcolor{myblue}{0.0208} & \cellcolor{myblue}{0.0119} & \cellcolor{myblue}{0.0317} & \cellcolor{myblue}{0.0152} & \cellcolor{myblue}{0.1178} & 0.0745 & \cellcolor{myblue}{0.1608} & \cellcolor{myblue}{0.0844} & \cellcolor{myblue}{0.0397} & \cellcolor{myblue}{0.0228} & \cellcolor{myblue}{0.0616} & \cellcolor{myblue}{0.0273} & -40.15\% \\
    \midrule
    BSARec \cite{shin2024attentive} & \multicolumn{1}{c|}{\multirow{2}[2]{*}{AAAI 2024}} & 0.0987  & 0.0595  & 0.1345  & 0.0685  & 0.0580  & 0.0338  & 0.0820  & 0.0399  & 0.1882  & 0.1199  & 0.2375  & 0.1323  & 0.0761  & 0.0441  & 0.1073  & 0.0519  & \multicolumn{1}{c}{—} \\
    w/o SSS &       & \cellcolor{myblue}{0.0569} & 0.0349  & \cellcolor{myblue}{0.0781} & 0.0402  & \cellcolor{myblue}{0.0233} & \cellcolor{myblue}{0.0143} & \cellcolor{myblue}{0.0333} & \cellcolor{myblue}{0.0168} & 0.1379  & 0.0875  & 0.1764  & 0.0972  & \cellcolor{myblue}{0.0220} & \cellcolor{myblue}{0.0109} & \cellcolor{myblue}{0.0257} & \cellcolor{myblue}{0.0118} & -50.46\% \\
    \midrule
    MUFFIN \cite{baek2025muffin} & \multicolumn{1}{c|}{\multirow{2}[2]{*}{CIKM 2025}} & 0.0912 & 0.0491 & 0.1283 & 0.0594 & 0.0568 & 0.0293 & 0.0829 & 0.0378 & 0.1780  & 0.1163 & 0.2203 & 0.1259 & 0.0759  & 0.0428 & 0.1105 & 0.0514 & \multicolumn{1}{c}{—} \\
    w/o SSS &       & \cellcolor{myblue}{0.0571} & 0.0343  & \cellcolor{myblue}{0.0874} & 0.0390  & \cellcolor{myblue}{0.0259} & \cellcolor{myblue}{0.0154} & \cellcolor{myblue}{0.0367} & \cellcolor{myblue}{0.0184} & 0.1423 & 0.0950  & 0.1855 & 0.1078 & \cellcolor{myblue}{0.0265} & \cellcolor{myblue}{0.0166} & \cellcolor{myblue}{0.0446} & \cellcolor{myblue}{0.0202} & -41.11\% \\
    \midrule
    CSRec \cite{liu2025csrec} & \multicolumn{1}{c|}{\multirow{2}[2]{*}{SIGIR 2025}} & 0.1016 & 0.0583 & 0.1372 & 0.0673 & 0.0565 & 0.0331 & 0.0789 & 0.0386  & 0.1836 & 0.1168 & 0.2311 & 0.1296 & 0.0792 & 0.0460  & 0.1096 & 0.0525 & \multicolumn{1}{c}{—} \\
    w/o SSS &       & 0.0602 & 0.0363 & \cellcolor{myblue}{0.0795} & 0.0413 & \cellcolor{myblue}{0.0246} & \cellcolor{myblue}{0.0147} & \cellcolor{myblue}{0.0349} & \cellcolor{myblue}{0.0174} & 0.1406 & 0.0869 & 0.1821 & 0.0975 & \cellcolor{myblue}{0.0269} & \cellcolor{myblue}{0.0129} & \cellcolor{myblue}{0.0314} & \cellcolor{myblue}{0.0152} & -47.33\% \\
    \midrule
    WaveRec \cite{heo2025waverec} & \multicolumn{1}{c|}{\multirow{2}[2]{*}{SIGIR 2025}} & 0.0871 & 0.0531 & 0.1172 & 0.0606 & 0.0524 & 0.0316 & 0.0708 & 0.0362 & 0.1502 & 0.0974 & 0.1854 & 0.1063 & \cellcolor{myblue}{0.0560} & \cellcolor{myblue}{0.0319} & \cellcolor{myblue}{0.0817} & \cellcolor{myblue}{0.0383} & \multicolumn{1}{c}{—} \\
    w/o SSS &       & \cellcolor{myblue}{0.0553} & 0.0334 & \cellcolor{myblue}{0.0746} & \cellcolor{myblue}{0.0382} & \cellcolor{myblue}{0.0260} & \cellcolor{myblue}{0.0153} & \cellcolor{myblue}{0.0365} & \cellcolor{myblue}{0.0179} & 0.1256 & 0.0816 & \cellcolor{myblue}{0.1574} & 0.0896 & \cellcolor{myblue}{0.0193} & \cellcolor{myblue}{0.0098} & \cellcolor{myblue}{0.0294} & \cellcolor{myblue}{0.0123} & -42.38\% \\
    \midrule
    TV-Rec \cite{shin2025tv} & \multicolumn{1}{c|}{\multirow{2}[2]{*}{NIPS 2025}} & 0.1017 & 0.0608 & 0.1403 & 0.0705 & 0.0594 & 0.0345 & 0.0859 & 0.0412 & 0.1964 & 0.1254 & 0.2461 & 0.1380  & 0.0853 & 0.0484 & 0.1202 & 0.0572 & \multicolumn{1}{c}{—} \\
    w/o SSS &       & 0.0602 & 0.0358 & \cellcolor{myblue}{0.0818} & 0.0412 & \cellcolor{myblue}{0.0271} & \cellcolor{myblue}{0.0159} & \cellcolor{myblue}{0.0381} & \cellcolor{myblue}{0.0187} & 0.1650  & 0.1012 & 0.2055 & 0.1114 & \cellcolor{myblue}{0.0193} & \cellcolor{myblue}{0.0073} & \cellcolor{myblue}{0.0275} & \cellcolor{myblue}{0.0094} & -48.61\% \\
    \midrule
    FreqRec \cite{he2025exploiting} & \multicolumn{1}{c|}{\multirow{2}[2]{*}{AAAI 2026}} & 0.0970  & 0.0590  & 0.1320  & 0.0678 & 0.0572 & 0.0336 & 0.0807 & 0.0396 & 0.1860  & 0.1174 & 0.2340  & 0.1296 & \cellcolor{myblue}{0.0615} & 0.0378 & \cellcolor{myblue}{0.0963} & 0.0466 & \multicolumn{1}{c}{—} \\
    w/o SSS &       & \cellcolor{myblue}{0.0532} & 0.0324 & \cellcolor{myblue}{0.0738} & \cellcolor{myblue}{0.0375} & \cellcolor{myblue}{0.0227} & \cellcolor{myblue}{0.0137} & \cellcolor{myblue}{0.0326} & \cellcolor{myblue}{0.0162} & 0.1502 & 0.0941 & 0.1910  & 0.1044 & \cellcolor{myblue}{0.0183} & \cellcolor{myblue}{0.0089} & \cellcolor{myblue}{0.0303} & \cellcolor{myblue}{0.0120} & -48.98\% \\
    \midrule
    WEARec \cite{xu2025wavelet} & \multicolumn{1}{c|}{\multirow{2}[2]{*}{AAAI 2026}} & 0.1000  & 0.0591 & 0.1366 & 0.0683 & 0.0612 & 0.0355 & 0.0876 & 0.0421 & 0.1919 & 0.1221 & 0.2387 & 0.1340  & 0.0761 & 0.0415 & 0.1055 & 0.0490  & \multicolumn{1}{c}{—} \\
    w/o SSS &       & 0.0614 & 0.0368 & \cellcolor{myblue}{0.0826} & 0.0421 & \cellcolor{myblue}{0.0302} & \cellcolor{myblue}{0.0176} & \cellcolor{myblue}{0.0424} & \cellcolor{myblue}{0.0207} & 0.1623 & 0.1010  & 0.2006 & 0.1107 & \cellcolor{myblue}{0.0147} & \cellcolor{myblue}{0.0078} & \cellcolor{myblue}{0.0321} & \cellcolor{myblue}{0.0122} & -45.65\% \\
    \bottomrule
    \end{tabular}}}%
  \label{tab:baseline_results}%
      \vspace{-0.5em}
\end{table*}%

\begin{table*}[!t]
  \centering
  \caption{Overall percentage of the recent SR models in Table \ref{tab:baseline_results} compared to SASRec before and after removing SSS.}
  \vspace{-1em}
      \renewcommand\arraystretch{0.9}
      \scalebox{0.98}{
    \begin{tabular}{c|cc|cc|cc|cc|cc}
    \toprule
    Model & \multicolumn{2}{c|}{FMLPRec \cite{zhou2022filter}} & \multicolumn{2}{c|}{DLFSRec \cite{liu2023distribution}} & \multicolumn{2}{c|}{DiffuRec \cite{li2023diffurec}} & \multicolumn{2}{c|}{BSARec \cite{shin2024attentive}} & \multicolumn{2}{c}{MUFFIN \cite{baek2025muffin}} \\
    Variant & Original & w/o SSS & Original & w/o SSS & Original & w/o SSS & Original & w/o SSS & Original & w/o SSS \\
    Comparison with SASRec & 0.13\% & 3.28\% & 16.37\% & 0.69\% & 23.09\% & -26.34\% & 55.85\% & -20.76\% & 46.64\% & -27.29\% \\
    \midrule
    Model & \multicolumn{2}{c|}{CSRec \cite{liu2025csrec}} & \multicolumn{2}{c|}{WaveRec \cite{heo2025waverec}} & \multicolumn{2}{c|}{TV-Rec \cite{shin2025tv}} & \multicolumn{2}{c|}{FreqRec \cite{he2025exploiting}} & \multicolumn{2}{c}{WEARec \cite{xu2025wavelet}} \\
    Variant & Original & w/o SSS & Original & w/o SSS & Original & w/o SSS & Original & w/o SSS & Original & w/o SSS \\
    Comparison with SASRec & 54.54\% & -17.07\% & 31.09\% & -22.75\% & 63.40\% & -13.85\% & 50.16\% & -21.13\% & 96.61\% & -11.04\% \\
    \bottomrule
    \end{tabular}}%
  \label{tab:baseline_results_percent}%
  \vspace{-1em}
\end{table*}%

\subsection{Experimental Settings}
\noindent \textbf{Datasets.} The experiments are conducted on four widely used datasets from different scenarios. \textbf{Beauty} and \textbf{Sports} are obtained from the Amazon platform \cite{mcauley2015inferring}, corresponding to the ``Beauty'' and ``Sports and Outdoors'' categories. \textbf{Douyin} \cite{zhang2024ninerec} includes short-video watching records. \textbf{LastFM}\footnote{https://grouplens.org/datasets/hetrec-2011/} includes artist listening records and is used to recommend musicians to users. We reduce the data by extracting the 5-core \cite{zhou2022filter,shin2024attentive,shin2025tv}. Following the setup of most of the investigated papers, the maximum sequence length is set to 50 for all datasets. The statistics are summarized in Table \ref{tab:datasets}.

\vspace{0.3em}

\noindent \textbf{Other Baselines.} In addition to the model mentioned earlier, we selected four representative SR models based on different architectures, including \textbf{GRU4Rec} (based on recurrent neural networks) \cite{hidasi2015session}, \textbf{SASRec} \cite{kang2018self} (based on Transformer), \textbf{NextItNet} (based on convolutional neural networks) \cite{yuan2019simple}, and \textbf{LRURec} (based on linear recurrent units) \cite{yue2024linear}. These models did not utilize SSS. Furthermore, for comparative purposes, we selected four SR models that incorporate contrastive learning, including \textbf{CLS4Rec} \cite{xie2022contrastive}, \textbf{CoSeRec} \cite{liu2021contrastive}, \textbf{DuoRec} \cite{qiu2022contrastive}, and \textbf{ICSRec} \cite{qin2024intent}. These methods typically employ data augmentation to generate self-supervised signals and use contrastive objectives to learn user preferences further.

\vspace{0.3em}

\noindent \textbf{Evaluation Settings.} We adopt the leave-one-out strategy to partition sequences into training, validation, and test sets. We rank predictions across the entire item set rather than using negative sampling, which can lead to biased discoveries \cite{krichene2020sampled}. The evaluation metrics include Hit Ratio@K (H@K) and Normalized Discounted Cumulative Gain@K (N@K). We report results with K $\in \{10, 20\}$. Generally, \emph{greater} values imply \emph{better} ranking accuracy. We run five trials and report the average results for all methods. 

\vspace{0.3em}

\noindent \textbf{Implementation Details.} For all baselines, we adopt the implementation provided by the authors. We set the embedding size to 64 and the batch size to 256. To ensure fair comparisons, we carefully set and tune all other hyperparameters of each method as reported and suggested in the original papers. We use the Adam \cite{2014Adam} optimizer with the learning rate 0.001, $\beta_1=0.9$, $\beta_2=0.999$. We adopt early stopping on the validation set if the performance does not improve for 20 epochs, and report results on test set \cite{liu2021contrastive,wang2026eeo}. 

\subsection{Results Analysis}

We present the performance of baselines and our investigated models in Table \ref{tab:baseline_results}. We give the performance before and after removing the SSS. We have the following observations and analysis:

(1) Among the four backbones without SSS, SASRec achieved optimal performance in almost all cases, demonstrating the Transformer's ability to model sequences. Contrastive learning enhances recommendation accuracy by extracting user preference knowledge from self-supervised signals. Some contrastive learning-based methods, such as DuoRec and ICSRec, achieve better performance by leveraging SSS to generate richer self-supervised signals.

(2) For SR models incorporating SSS, we observe that most methods achieve performance significantly better than that of contrastive learning-based SR models (and the SR models without SSS). This result reaffirms the conclusions drawn in prior work \cite{zhou2024contrastive}. However, when we removed SSS from the data reading phase of these models, their performance declined significantly. Eight out of ten models show an average performance drop exceeding 40\%, with five models experiencing nearly a 50\% reduction. Many models proposed in the past year or two now perform worse than SASRec (highlighted in blue), which was introduced back in 2018.

(3) To further illustrate how SSS impacts researchers' evaluation of the actual performance of these latest models, Table \ref{tab:baseline_results_percent} presents the overall performance percentage relative to SASRec for these models before and after SSS removal. We can observe that after removing SSS, nearly all models no longer outperform SASRec. It bears repeating that the original SASRec did not employ sub-sequence splitting. Researchers may have mistakenly attributed the remarkable performance gains from SSS to improvements in model architecture. Additionally, we observed that the number of models influenced by SSS increased in publications from 2023 to 2025, with the vast majority proposed in 2025. Considering that conference and journal publications for 2026 have only just begun, more SSS-influenced models will likely be published in 2026. Therefore, attention to this issue is urgently needed.

\begin{table*}[!t]
  \centering
  \caption{The performance of four backbone SR models with different training settings on four small-scale datasets. The best performance in each case is highlighted in bolded \textcolor{textred}{red} and the second-best is highlighted in underlined \textcolor{textblue}{blue}.}
      \vspace{-1em}
    \renewcommand\arraystretch{0.875}
    \setlength{\tabcolsep}{1.0mm}{
      \scalebox{0.75}{
    \begin{tabular}{c|c|c|c||cccc|cccc|cccc|cccc}
    \toprule
    \multirow{2}[2]{*}{Model} & \multirow{2}[2]{*}{Target} & \multirow{2}[2]{*}{Loss} & \multirow{2}[2]{*}{Split} & \multicolumn{4}{c|}{Beauty}   & \multicolumn{4}{c|}{Sports}   & \multicolumn{4}{c|}{Douyin}   & \multicolumn{4}{c}{LastFM} \\
          &       &       &       & H@10  & N@10  & H@20  & N@20  & H@10  & N@10  & H@20  & N@20  & H@10  & N@10  & H@20  & N@20  & H@10  & N@10  & H@20  & N@20 \\
    \midrule
    \midrule
    \multirow{16}[8]{*}{GRU4Rec} & \multirow{8}[4]{*}{Single} & \multirow{4}[2]{*}{BCE} & Prefix & 0.0312  & 0.0153  & 0.0493  & 0.0198  & 0.0207  & 0.0110  & 0.0319  & 0.0138  & 0.0689  & 0.0347  & 0.1072  & 0.0443  & 0.0321  & 0.0177  & 0.0541  & 0.0231  \\
          &       &       & Suffix & 0.0250  & 0.0125  & 0.0419  & 0.0167  & 0.0112  & 0.0054  & 0.0188  & 0.0073  & 0.0578  & 0.0302  & 0.0914  & 0.0387  & 0.0220  & 0.0157  & 0.0303  & 0.0178  \\
          &       &       & Slide & 0.0236  & 0.0127  & 0.0412  & 0.0171  & 0.0162  & 0.0083  & 0.0270  & 0.0110  & 0.0822  & 0.0433  & 0.1203  & 0.0529  & 0.0339  & 0.0185  & 0.0514  & 0.0231  \\
          &       &       & Original & 0.0152  & 0.0075  & 0.0250  & 0.0100  & 0.0097  & 0.0056  & 0.0173  & 0.0075  & 0.0258  & 0.0125  & 0.0442  & 0.0171  & 0.0092  & 0.0050  & 0.0220  & 0.0082  \\
\cmidrule{3-20}          &       & \multirow{4}[2]{*}{CE} & Prefix & \cellcolor{myblue} \underline{0.0688}  & \cellcolor{myblue} \underline{0.0379}  & \cellcolor{myblue} \underline{0.0999}  & \cellcolor{myblue} \underline{0.0458}  & \cellcolor{myred} \textbf{0.0401} & \cellcolor{myred} \textbf{0.0215} & \cellcolor{myred} \textbf{0.0605} & \cellcolor{myred} \textbf{0.0266} & \cellcolor{myred} \textbf{0.1414} & \cellcolor{myred} \textbf{0.0878} & \cellcolor{myred} \textbf{0.1839} & \cellcolor{myred} \textbf{0.0986} & \cellcolor{myred} \textbf{0.0606} & \cellcolor{myred} \textbf{0.0315} & 0.0807  & \cellcolor{myblue} \underline{0.0365}  \\
          &       &       & Suffix & 0.0343  & 0.0198  & 0.0494  & 0.0236  & 0.0135  & 0.0071  & 0.0216  & 0.0092  & 0.0966  & 0.0621  & 0.1219  & 0.0685  & 0.0193  & 0.0135  & 0.0275  & 0.0157  \\
          &       &       & Slide & 0.0616  & 0.0348  & 0.0881  & 0.0414  & \cellcolor{myblue} \underline{0.0336}  & \cellcolor{myblue} \underline{0.0181}  & \cellcolor{myblue} \underline{0.0512}  & \cellcolor{myblue} \underline{0.0225}  & 0.1271  & 0.0780  & 0.1654  & \cellcolor{myblue} \underline{0.0876}  & 0.0459  & 0.0235  & 0.0725  & 0.0304  \\
          &       &       & Original & 0.0330  & 0.0179  & 0.0485  & 0.0217  & 0.0183  & 0.0092  & 0.0288  & 0.0119  & 0.1023  & 0.0598  & 0.1322  & 0.0673  & 0.0257  & 0.0126  & 0.0450  & 0.0175  \\
\cmidrule{2-20}          & \multirow{8}[4]{*}{Multi} & \multirow{4}[2]{*}{BCE} & Prefix & 0.0444  & 0.0230  & 0.0706  & 0.0296  & 0.0169  & 0.0082  & 0.0285  & 0.0111  & 0.1079  & 0.0591  & 0.1482  & 0.0693  & 0.0560  & \cellcolor{myblue} \underline{0.0296}  & \cellcolor{myred} \textbf{0.0862} & \cellcolor{myred} \textbf{0.0372} \\
          &       &       & Suffix & 0.0576  & 0.0306  & 0.0841  & 0.0373  & 0.0253  & 0.0129  & 0.0409  & 0.0168  & \cellcolor{myblue} \underline{0.1333}  & \cellcolor{myblue} \underline{0.0788}  & \cellcolor{myblue} \underline{0.1680}  & \cellcolor{myblue} \underline{0.0876}  & 0.0495  & 0.0260  & 0.0725  & 0.0318  \\
          &       &       & Slide & 0.0528  & 0.0287  & 0.0804  & 0.0356  & 0.0220  & 0.0116  & 0.0358  & 0.0150  & 0.1235  & 0.0696  & 0.1627  & 0.0795  & \cellcolor{myblue} \underline{0.0569}  & 0.0291  & \cellcolor{myblue} \underline{0.0853}  & 0.0362  \\
          &       &       & Original & 0.0423  & 0.0209  & 0.0704  & 0.0279  & 0.0180  & 0.0085  & 0.0311  & 0.0118  & 0.1060  & 0.0572  & 0.1480  & 0.0678  & 0.0385  & 0.0214  & 0.0569  & 0.0260  \\
\cmidrule{3-20}          &       & \multirow{4}[2]{*}{CE} & Prefix & 0.0442  & 0.0251  & 0.0639  & 0.0300  & 0.0193  & 0.0109  & 0.0305  & 0.0138  & 0.0974  & 0.0608  & 0.1274  & 0.0684  & 0.0239  & 0.0115  & 0.0514  & 0.0184  \\
          &       &       & Suffix & 0.0526  & 0.0311  & 0.0723  & 0.0361  & 0.0228  & 0.0132  & 0.0352  & 0.0163  & 0.1222  & 0.0785  & 0.1534  & 0.0863  & 0.0339  & 0.0174  & 0.0596  & 0.0240  \\
          &       &       & Slide & \cellcolor{myred} \textbf{0.0775} & \cellcolor{myred} \textbf{0.0470} & \cellcolor{myred} \textbf{0.1054} & \cellcolor{myred} \textbf{0.0540} & 0.0292  & 0.0166  & 0.0432  & 0.0201  & 0.1108  & 0.0689  & 0.1419  & 0.0768  & 0.0486  & 0.0259  & 0.0688  & 0.0309  \\
          &       &       & Original & 0.0508  & 0.0290  & 0.0718  & 0.0343  & 0.0206  & 0.0106  & 0.0321  & 0.0135  & 0.1143  & 0.0722  & 0.1462  & 0.0803  & 0.0495  & 0.0258  & 0.0761  & 0.0324  \\
    \midrule
    \midrule
    \multirow{16}[7]{*}{SASRec} & \multirow{8}[4]{*}{Single} & \multirow{4}[2]{*}{BCE} & Prefix & 0.0455  & 0.0240  & 0.0735  & 0.0310  & 0.0282  & 0.0142  & 0.0477  & 0.0183  & 0.1095  & 0.0617  & 0.1542  & 0.0730  & 0.0606  & 0.0334  & 0.0872  & 0.0401  \\
          &       &       & Suffix & 0.0471  & 0.0254  & 0.0673  & 0.0305  & 0.0203  & 0.0105  & 0.0296  & 0.0129  & 0.1052  & 0.0628  & 0.1378  & 0.0710  & 0.0138  & 0.0063  & 0.0174  & 0.0072  \\
          &       &       & Slide & 0.0529  & 0.0273  & 0.0789  & 0.0338  & 0.0280  & 0.0150  & 0.0427  & 0.0187  & 0.1078  & 0.0638  & 0.1484  & 0.0740  & 0.0450  & 0.0249  & 0.0688  & 0.0308  \\
          &       &       & Original & 0.0463  & 0.0259  & 0.0657  & 0.0308  & 0.0100  & 0.0058  & 0.0162  & 0.0073  & 0.1002  & 0.0604  & 0.1291  & 0.0677  & 0.0138  & 0.0064  & 0.0239  & 0.0089  \\
\cmidrule{3-20}          &       & \multirow{4}[2]{*}{CE} & Prefix & \cellcolor{myblue} \underline{0.0944}  & \cellcolor{myred} \textbf{0.0572} & \cellcolor{myblue} \underline{0.1270}  & \cellcolor{myblue} \underline{0.0654}  & \cellcolor{myred} \textbf{0.0550} & \cellcolor{myred} \textbf{0.0323} & \cellcolor{myred} \textbf{0.0777} & \cellcolor{myred} \textbf{0.0380} & \cellcolor{myblue} \underline{0.1622}  & \cellcolor{myblue} \underline{0.1032}  & \cellcolor{myblue} \underline{0.2030}  & \cellcolor{myblue} \underline{0.1135}  & \cellcolor{myred} \textbf{0.0661} & \cellcolor{myred} \textbf{0.0395} & 0.0862  & \cellcolor{myred} \textbf{0.0447} \\
          &       &       & Suffix & 0.0464  & 0.0270  & 0.0692  & 0.0327  & 0.0195  & 0.0107  & 0.0279  & 0.0129  & 0.1256  & 0.0807  & 0.1579  & 0.0889  & 0.0183  & 0.0087  & 0.0229  & 0.0099  \\
          &       &       & Slide & 0.0815  & 0.0500  & 0.1122  & 0.0577  & 0.0456  & 0.0266  & 0.0640  & 0.0313  & 0.1583  & 0.1006  & 0.2000  & 0.1111  & 0.0514  & 0.0305  & 0.0734  & 0.0360  \\
          &       &       & Original & 0.0537  & 0.0321  & 0.0722  & 0.0368  & 0.0242  & 0.0145  & 0.0334  & 0.0168  & 0.1414  & 0.0901  & 0.1766  & 0.0990  & 0.0110  & 0.0047  & 0.0257  & 0.0084  \\
\cmidrule{2-20}          & \multirow{8}[3]{*}{Multi} & \multirow{4}[2]{*}{BCE} & Prefix & 0.0594  & 0.0316  & 0.0868  & 0.0385  & 0.0301  & 0.0162  & 0.0459  & 0.0202  & 0.1181  & 0.0664  & 0.1608  & 0.0771  & \cellcolor{myblue} \underline{0.0615}  & \cellcolor{myblue} \underline{0.0365}  & 0.0872  & \cellcolor{myblue} \underline{0.0430} \\
          &       &       & Suffix & 0.0674  & 0.0361  & 0.0968  & 0.0434  & 0.0303  & 0.0162  & 0.0479  & 0.0206  & 0.1277  & 0.0747  & 0.1686  & 0.0850  & 0.0440  & 0.0255  & 0.0771  & 0.0339  \\
          &       &       & Slide & 0.0596  & 0.0311  & 0.0863  & 0.0378  & 0.0278  & 0.0151  & 0.0427  & 0.0188  & 0.1254  & 0.0735  & 0.1689  & 0.0845  & 0.0358  & 0.0212  & 0.0569  & 0.0265  \\
          &       &       & Original & 0.0580  & 0.0301  & 0.0910  & 0.0384  & 0.0323  & 0.0179  & 0.0490  & 0.0222  & 0.1243  & 0.0739  & 0.1677  & 0.0848  & \cellcolor{myred} \textbf{0.0661} & 0.0347  & \cellcolor{myred} \textbf{0.0982} & 0.0428  \\
\cmidrule{3-20}          &       & \multirow{4}[1]{*}{CE} & Prefix & 0.0731  & 0.0444  & 0.1002  & 0.0512  & 0.0385  & 0.0224  & 0.0558  & 0.0267  & 0.1562  & 0.0971  & 0.1987  & 0.1078  & 0.0596  & 0.0342  & 0.0780  & 0.0389  \\
          &       &       & Suffix & 0.0727  & 0.0424  & 0.1011  & 0.0496  & 0.0382  & 0.0213  & 0.0548  & 0.0255  & 0.1543  & 0.0975  & 0.1954  & 0.1078  & 0.0468  & 0.0231  & 0.0734  & 0.0298  \\
          &       &       & Slide & 0.0804  & 0.0493  & 0.1082  & 0.0564  & 0.0442  & 0.0263  & 0.0620  & 0.0308  & 0.1572  & 0.0975  & 0.2023  & 0.1089  & 0.0459  & 0.0290  & 0.0761  & 0.0366  \\
          &       &       & Original & \cellcolor{myred} \textbf{0.0947} & \cellcolor{myblue} \underline{0.0569}  & \cellcolor{myred} \textbf{0.1312} & \cellcolor{myred} \textbf{0.0661} & \cellcolor{myblue} \underline{0.0527}  & \cellcolor{myblue} \underline{0.0306}  & \cellcolor{myblue} \underline{0.0755}  & \cellcolor{myblue} \underline{0.0363}  & \cellcolor{myred} \textbf{0.1718} & \cellcolor{myred} \textbf{0.1094} & \cellcolor{myred} \textbf{0.2146} & \cellcolor{myred} \textbf{0.1202} & \cellcolor{myblue} \underline{0.0615}  & 0.0338  & \cellcolor{myblue} \underline{0.0927}  & 0.0416  \\
    \midrule
    \midrule
    \multirow{16}[7]{*}{FMLPRec} & \multirow{8}[3]{*}{Single} & \multirow{4}[1]{*}{BCE} & Prefix & 0.0564  & 0.0292  & 0.0867  & 0.0368  & 0.0314  & 0.0171  & 0.0505  & 0.0219  & 0.1272  & 0.0716  & 0.1819  & 0.0853  & \cellcolor{myred} \textbf{0.0670} & 0.0324  & \cellcolor{myred} \textbf{0.1000} & 0.0408  \\
          &       &       & Suffix & 0.0510  & 0.0284  & 0.0724  & 0.0338  & 0.0220  & 0.0114  & 0.0332  & 0.0142  & 0.1142  & 0.0669  & 0.1489  & 0.0757  & 0.0083  & 0.0042  & 0.0202  & 0.0072  \\
          &       &       & Slide & 0.0554  & 0.0291  & 0.0848  & 0.0365  & 0.0316  & 0.0173  & 0.0484  & 0.0215  & 0.1261  & 0.0703  & 0.1762  & 0.0830  & 0.0477  & 0.0302  & 0.0752  & 0.0371  \\
          &       &       & Original & 0.0446  & 0.0250  & 0.0632  & 0.0296  & 0.0096  & 0.0054  & 0.0201  & 0.0080  & 0.1095  & 0.0642  & 0.1417  & 0.0723  & 0.0147  & 0.0070  & 0.0220  & 0.0089  \\
\cmidrule{3-20}          &       & \multirow{4}[2]{*}{CE} & Prefix & \cellcolor{myred} \textbf{0.0962} & \cellcolor{myred} \textbf{0.0568} & \cellcolor{myred} \textbf{0.1324} & \cellcolor{myred} \textbf{0.0660} & \cellcolor{myred} \textbf{0.0706} & \cellcolor{myred} \textbf{0.0401} & \cellcolor{myred} \textbf{0.1128} & \cellcolor{myred} \textbf{0.0507} & \cellcolor{myred} \textbf{0.1957} & \cellcolor{myred} \textbf{0.1243} & \cellcolor{myred} \textbf{0.2452} & \cellcolor{myred} \textbf{0.1369} & 0.0584  & 0.0333  & 0.0836  & 0.0397  \\
          &       &       & Suffix & 0.0513  & 0.0297  & 0.0718  & 0.0348  & 0.0183  & 0.0080  & 0.0229  & 0.0092  & 0.1494  & 0.0934  & 0.1854  & 0.1026  & 0.0223  & 0.0124  & 0.0322  & 0.0149  \\
          &       &       & Slide & 0.0894  & 0.0541  & 0.1236  & 0.0627  & \cellcolor{myblue} \underline{0.0615}  & \cellcolor{myblue} \underline{0.0340}  & \cellcolor{myblue} \underline{0.0927}  & \cellcolor{myblue} \underline{0.0417}  & 0.1779  & 0.1114  & 0.2217  & 0.1224  & 0.0513  & 0.0293  & 0.0719  & 0.0344  \\
          &       &       & Original & 0.0557  & 0.0345  & 0.0753  & 0.0395  & 0.0119  & 0.0049  & 0.0257  & 0.0082  & 0.1592  & 0.0997  & 0.1976  & 0.1094  & 0.0250  & 0.0149  & 0.0349  & 0.0174  \\
\cmidrule{2-20}          & \multirow{8}[4]{*}{Multi} & \multirow{4}[2]{*}{BCE} & Prefix & 0.0620  & 0.0331  & 0.0908  & 0.0403  & 0.0309  & 0.0161  & 0.0474  & 0.0203  & 0.1245  & 0.0706  & 0.1714  & 0.0825  & 0.0633  & \cellcolor{myred} \textbf{0.0375} & \cellcolor{myblue} \underline{0.0982}  & \cellcolor{myred} \textbf{0.0461} \\
          &       &       & Suffix & 0.0658  & 0.0359  & 0.0974  & 0.0439  & 0.0309  & 0.0170  & 0.0478  & 0.0213  & 0.1301  & 0.0763  & 0.1709  & 0.0866  & 0.0532  & 0.0281  & 0.0835  & 0.0356  \\
          &       &       & Slide & 0.0566  & 0.0304  & 0.0846  & 0.0374  & 0.0288  & 0.0154  & 0.0441  & 0.0192  & 0.1268  & 0.0730  & 0.1724  & 0.0845  & 0.0459  & 0.0255  & 0.0633  & 0.0299  \\
          &       &       & Original & 0.0607  & 0.0331  & 0.0926  & 0.0411  & 0.0326  & 0.0175  & 0.0507  & 0.0221  & 0.1316  & 0.0758  & 0.1796  & 0.0879  & \cellcolor{myblue} \underline{0.0642}  & 0.0366  & \cellcolor{myblue} \underline{0.0982}  & \cellcolor{myblue} \underline{0.0452}  \\
\cmidrule{3-20}          &       & \multirow{4}[2]{*}{CE} & Prefix & 0.0722  & 0.0440  & 0.1000  & 0.0510  & 0.0366  & 0.0212  & 0.0543  & 0.0256  & 0.1555  & 0.0979  & 0.1964  & 0.1082  & 0.0606  & 0.0355  & 0.0853  & 0.0418  \\
          &       &       & Suffix & 0.0737  & 0.0438  & 0.1020  & 0.0509  & 0.0345  & 0.0194  & 0.0513  & 0.0236  & 0.1525  & 0.0969  & 0.1947  & 0.1075  & 0.0422  & 0.0232  & 0.0743  & 0.0313  \\
          &       &       & Slide & 0.0811  & 0.0502  & 0.1086  & 0.0571  & 0.0444  & 0.0266  & 0.0632  & 0.0313  & 0.1583  & 0.0986  & 0.2006  & 0.1093  & 0.0459  & 0.0294  & 0.0651  & 0.0343  \\
          &       &       & Original & \cellcolor{myblue} \underline{0.0930}  & \cellcolor{myblue} \underline{0.0566}  & \cellcolor{myblue} \underline{0.1273}  & \cellcolor{myblue} \underline{0.0652}  & 0.0511  & 0.0299  & 0.0737  & 0.0356  & \cellcolor{myblue} \underline{0.1790}  & \cellcolor{myblue} \underline{0.1133}  & \cellcolor{myblue} \underline{0.2272}  & \cellcolor{myblue} \underline{0.1255}  & \cellcolor{myred} \textbf{0.0670} & \cellcolor{myblue} \underline{0.0371}  & 0.0936  & 0.0437  \\
    \midrule
    \midrule
    \multirow{16}[8]{*}{BSARec} & \multirow{8}[4]{*}{Single} & \multirow{4}[2]{*}{BCE} & Prefix & 0.0578  & 0.0304  & 0.0872  & 0.0378  & 0.0295  & 0.0158  & 0.0464  & 0.0200  & 0.1289  & 0.0707  & 0.1832  & 0.0843  & 0.0642  & 0.0397  & 0.0972  & 0.0481  \\
          &       &       & Suffix & 0.0503  & 0.0286  & 0.0715  & 0.0339  & 0.0208  & 0.0107  & 0.0305  & 0.0131  & 0.1143  & 0.0688  & 0.1495  & 0.0777  & 0.0138  & 0.0078  & 0.0239  & 0.0103  \\
          &       &       & Slide & 0.0609  & 0.0327  & 0.0887  & 0.0397  & 0.0304  & 0.0162  & 0.0471  & 0.0204  & 0.1228  & 0.0706  & 0.1661  & 0.0816  & 0.0523  & 0.0279  & 0.0743  & 0.0332  \\
          &       &       & Original & 0.0489  & 0.0273  & 0.0690  & 0.0324  & 0.0127  & 0.0065  & 0.0193  & 0.0081  & 0.1032  & 0.0618  & 0.1365  & 0.0702  & 0.0092  & 0.0036  & 0.0294  & 0.0086  \\
\cmidrule{3-20}          &       & \multirow{4}[2]{*}{CE} & Prefix & \cellcolor{myred} \textbf{0.0987} & \cellcolor{myred} \textbf{0.0595} & \cellcolor{myred} \textbf{0.1345} & \cellcolor{myred} \textbf{0.0685} & \cellcolor{myred} \textbf{0.0580} & \cellcolor{myred} \textbf{0.0338} & \cellcolor{myred} \textbf{0.0820} & \cellcolor{myred} \textbf{0.0399} & \cellcolor{myred} \textbf{0.1882} & \cellcolor{myred} \textbf{0.1199} & \cellcolor{myred} \textbf{0.2375} & \cellcolor{myred} \textbf{0.1323} & 0.0761  & 0.0441  & 0.1073  & 0.0519  \\
          &       &       & Suffix & 0.0509  & 0.0290  & 0.0709  & 0.0341  & 0.0217  & 0.0120  & 0.0317  & 0.0146  & 0.1379  & 0.0875  & 0.1764  & 0.0972  & 0.0220  & 0.0109  & 0.0257  & 0.0118  \\
          &       &       & Slide & \cellcolor{myblue} \underline{0.0868}  & \cellcolor{myblue} \underline{0.0525}  & 0.1170  & \cellcolor{myblue} \underline{0.0602}  & 0.0461  & \cellcolor{myblue} \underline{0.0275}  & 0.0655  & \cellcolor{myblue} \underline{0.0324}  & \cellcolor{myblue} \underline{0.1795}  & \cellcolor{myblue} \underline{0.1144}  & \cellcolor{myblue} \underline{0.2250}  & \cellcolor{myblue} \underline{0.1258}  & 0.0596  & 0.0341  & 0.0963  & 0.0434  \\
          &       &       & Original & 0.0569  & 0.0349  & 0.0781  & 0.0402  & 0.0233  & 0.0143  & 0.0333  & 0.0168  & 0.1569  & 0.0981  & 0.1996  & 0.1088  & 0.0211  & 0.0104  & 0.0349  & 0.0139  \\
\cmidrule{2-20}          & \multirow{8}[4]{*}{Multi} & \multirow{4}[2]{*}{BCE} & Prefix & 0.0732  & 0.0402  & 0.1069  & 0.0487  & 0.0372  & 0.0199  & 0.0574  & 0.0250  & 0.1363  & 0.0789  & 0.1866  & 0.0916  & \cellcolor{myred} \textbf{0.0826} & \cellcolor{myred} \textbf{0.0459} & \cellcolor{myred} \textbf{0.1147} & \cellcolor{myred} \textbf{0.0539} \\
          &       &       & Suffix & 0.0745  & 0.0425  & 0.1057  & 0.0503  & 0.0346  & 0.0188  & 0.0535  & 0.0236  & 0.1419  & 0.0845  & 0.1843  & 0.0951  & 0.0615  & 0.0304  & 0.0982  & 0.0395  \\
          &       &       & Slide & 0.0699  & 0.0387  & 0.0986  & 0.0459  & 0.0340  & 0.0183  & 0.0513  & 0.0227  & 0.1404  & 0.0813  & 0.1896  & 0.0937  & 0.0606  & 0.0333  & 0.0899  & 0.0407  \\
          &       &       & Original & 0.0713  & 0.0386  & 0.1043  & 0.0469  & 0.0353  & 0.0190  & 0.0544  & 0.0238  & 0.1378  & 0.0795  & 0.1871  & 0.0920  & 0.0725  & 0.0418  & \cellcolor{myblue} \underline{0.1101}  & 0.0511  \\
\cmidrule{3-20}          &       & \multirow{4}[2]{*}{CE} & Prefix & 0.0741  & 0.0451  & 0.1009  & 0.0518  & 0.0386  & 0.0222  & 0.0556  & 0.0265  & 0.1641  & 0.1023  & 0.2082  & 0.1135  & 0.0679  & 0.0366  & 0.0927  & 0.0428  \\
          &       &       & Suffix & 0.0703  & 0.0422  & 0.0990  & 0.0495  & 0.0366  & 0.0203  & 0.0533  & 0.0245  & 0.1648  & 0.1032  & 0.2094  & 0.1145  & 0.0358  & 0.0187  & 0.0578  & 0.0242  \\
          &       &       & Slide & 0.0811  & 0.0500  & 0.1101  & 0.0572  & 0.0430  & 0.0257  & 0.0599  & 0.0300  & 0.1607  & 0.1016  & 0.2030  & 0.1123  & 0.0578  & 0.0341  & 0.0817  & 0.0401  \\
          &       &       & Original & 0.0827  & 0.0503  & \cellcolor{myblue} \underline{0.1171}  & 0.0589  & \cellcolor{myblue} \underline{0.0474}  & 0.0268  & \cellcolor{myblue} \underline{0.0689}  & 0.0322  & 0.1683  & 0.1070  & 0.2107  & 0.1177  & \cellcolor{myblue} \underline{0.0771}  & \cellcolor{myblue} \underline{0.0444}  & 0.1073  & \cellcolor{myblue} \underline{0.0520}  \\
    \bottomrule
    \end{tabular}}}%
    \vspace{-1em}
  \label{tab:when_effective_1}%
\end{table*}%

\section{When is SSS Effective}\label{sec:when}
In the previous section, we demonstrated how SSS interferes with researchers' evaluation of a model's actual performance. Beyond the model architecture itself, different models may employ distinct loss functions, splitting methods (introduced in Subsection \ref{sec:sss}), and target strategies (introduced in Subsection \ref{sec:target}). We conducted comprehensive experiments to reveal how SSS interacts with these factors to jointly influence model performance.

\begin{table*}[!t]
  \centering
  \caption{The performance of four backbone SR models with different training settings on two large-scale datasets. The best performance in each case is highlighted in bolded \textcolor{textred}{red} and the second-best is highlighted in underlined \textcolor{textblue}{blue}.}
    \vspace{-1em}
    \renewcommand\arraystretch{0.875}
    \setlength{\tabcolsep}{1.2mm}{
      \scalebox{0.75}{
    \begin{tabular}{c|c|c||cccc|cccc||cccc|cccc}
    \toprule
    \multirow{2}[2]{*}{Target} & \multirow{2}[2]{*}{Loss} & \multirow{2}[2]{*}{Split} & \multicolumn{4}{c|}{ML-1M}    & \multicolumn{4}{c||}{CDs}     & \multicolumn{4}{c|}{ML-1M}    & \multicolumn{4}{c}{CDs} \\
          &       &       & H@10  & N@10  & H@20  & N@20  & H@10  & N@10  & H@20  & N@20  & H@10  & N@10  & H@20  & N@20  & H@10  & N@10  & H@20  & N@20 \\
    \midrule
    \midrule
    \multicolumn{3}{c||}{Model} & \multicolumn{8}{c||}{GRU4Rec}                                 & \multicolumn{8}{c}{SASRec} \\
    \midrule
    \multirow{8}[4]{*}{Single} & \multirow{4}[2]{*}{BCE} & Prefix & 0.1728  & 0.0872  & 0.2725  & 0.1123  & 0.0197  & 0.0092  & 0.0354  & 0.0131  & 0.2200  & 0.1181  & 0.3252  & 0.1445  & 0.0347  & 0.0174  & 0.0573  & 0.0231  \\
          &       & Suffix & 0.0614  & 0.0325  & 0.0974  & 0.0416  & 0.0038  & 0.0019  & 0.0072  & 0.0027  & 0.0902  & 0.0491  & 0.1272  & 0.0584  & 0.0200  & 0.0106  & 0.0297  & 0.0131  \\
          &       & Silde & 0.1689  & 0.0850  & 0.2654  & 0.1093  & 0.0138  & 0.0067  & 0.0247  & 0.0095  & 0.2136  & 0.1132  & 0.3101  & 0.1374  & 0.0362  & 0.0184  & 0.0564  & 0.0235  \\
          &       & Original & 0.0556  & 0.0280  & 0.0836  & 0.0351  & 0.0043  & 0.0021  & 0.0076  & 0.0029  & 0.1008  & 0.0567  & 0.1399  & 0.0665  & 0.0051  & 0.0023  & 0.0085  & 0.0032  \\
\cmidrule{2-19}          & \multirow{4}[2]{*}{CE} & Prefix & 0.2675  & 0.1494  & 0.3685  & 0.1749  & \cellcolor{myblue} \underline{0.0661} & \cellcolor{myblue} \underline{0.0352} & \cellcolor{myblue} \underline{0.0987} & \cellcolor{myblue} \underline{0.0434} & \cellcolor{myblue} \underline{0.2937} & \cellcolor{myblue} \underline{0.1697} & \cellcolor{myblue} \underline{0.3944} & \cellcolor{myblue} \underline{0.1951} & \cellcolor{myred} \textbf{0.0918} & \cellcolor{myred} \textbf{0.0515} & \cellcolor{myblue} \underline{0.1289} & \cellcolor{myred} \textbf{0.0609} \\
          &       & Suffix & 0.0608  & 0.0321  & 0.0937  & 0.0404  & 0.0142  & 0.0082  & 0.0210  & 0.0099  & 0.0848  & 0.0499  & 0.1185  & 0.0583  & 0.0221  & 0.0122  & 0.0317  & 0.0146  \\
          &       & Silde & 0.2492  & 0.1381  & 0.3538  & 0.1644  & 0.0518  & 0.0272  & 0.0790  & 0.0341  & 0.2853  & 0.1630  & 0.3811  & 0.1872  & 0.0749  & 0.0422  & 0.1064  & 0.0501  \\
          &       & Original & 0.0566  & 0.0276  & 0.0882  & 0.0357  & 0.0111  & 0.0057  & 0.0179  & 0.0074  & 0.0934  & 0.0539  & 0.1310  & 0.0634  & 0.0283  & 0.0158  & 0.0403  & 0.0188  \\
    \midrule
    \multirow{8}[4]{*}{Multi} & \multirow{4}[2]{*}{BCE} & Prefix & 0.2373  & 0.1221  & 0.3459  & 0.1494  & 0.0416  & 0.0210  & 0.0661  & 0.0271  & 0.2712  & 0.1475  & 0.3821  & 0.1755  & 0.0527  & 0.0227  & 0.0815  & 0.0349  \\
          &       & Suffix & 0.1755  & 0.0892  & 0.2616  & 0.1109  & 0.0441  & 0.0227  & 0.0710  & 0.0294  & 0.1719  & 0.0888  & 0.2765  & 0.1152  & 0.0501  & 0.0258  & 0.0783  & 0.0329  \\
          &       & Silde & 0.2550  & 0.1345  & 0.3719  & 0.1639  & 0.0341  & 0.0166  & 0.0562  & 0.0222  & 0.2353  & 0.1283  & 0.3376  & 0.1541  & 0.0402  & 0.0209  & 0.0640  & 0.0268  \\
          &       & Original & 0.1902  & 0.0967  & 0.2993  & 0.1241  & 0.0321  & 0.0156  & 0.0548  & 0.0213  & 0.2192  & 0.1167  & 0.3300  & 0.1447  & 0.0489  & 0.0253  & 0.0766  & 0.0323  \\
\cmidrule{2-19}          & \multirow{4}[2]{*}{CE} & Prefix & \cellcolor{myblue} \underline{0.2722} & \cellcolor{myblue} \underline{0.1523} & \cellcolor{myblue} \underline{0.3841} & \cellcolor{myblue} \underline{0.1805} & 0.0507  & 0.0270  & 0.0769  & 0.0336  & \cellcolor{myred} \textbf{0.3017} & \cellcolor{myred} \textbf{0.1710} & \cellcolor{myred} \textbf{0.4060} & \cellcolor{myred} \textbf{0.1972} & 0.0645  & 0.0348  & 0.0947  & 0.0424  \\
          &       & Suffix & 0.1538  & 0.0789  & 0.2356  & 0.0995  & 0.0445  & 0.0241  & 0.0660  & 0.0295  & 0.1942  & 0.1040  & 0.2838  & 0.1266  & 0.0457  & 0.0238  & 0.0695  & 0.0297  \\
          &       & Silde & \cellcolor{myred} \textbf{0.2942} & \cellcolor{myred} \textbf{0.1654} & \cellcolor{myred} \textbf{0.4109} & \cellcolor{myred} \textbf{0.1948} & \cellcolor{myred} \textbf{0.1108} & \cellcolor{myred} \textbf{0.0689} & \cellcolor{myred} \textbf{0.1419} & \cellcolor{myred} \textbf{0.0768} & 0.2783  & 0.1524  & 0.3886  & 0.1802  & 0.0733  & 0.0415  & 0.1030  & 0.0490  \\
          &       & Original & 0.2382  & 0.1292  & 0.3434  & 0.1557  & 0.0489  & 0.0260  & 0.0743  & 0.0324  & 0.2818  & 0.1567  & 0.3838  & 0.1824  & \cellcolor{myblue} \underline{0.0915} & \cellcolor{myblue} \underline{0.0512} & \cellcolor{myred} \textbf{0.1293} & \cellcolor{myblue} \underline{0.0607} \\
    \midrule
    \midrule
    \multicolumn{3}{c||}{Model} & \multicolumn{8}{c||}{FMLPRec}                                 & \multicolumn{8}{c}{BSARec} \\
    \midrule
    \multirow{8}[4]{*}{Single} & \multirow{4}[2]{*}{BCE} & Prefix & 0.2048  & 0.1057  & 0.3166  & 0.1338  & 0.0455  & 0.0230  & 0.0720  & 0.0296  & 0.2141  & 0.1115  & 0.3209  & 0.1383  & 0.0438  & 0.0225  & 0.0698  & 0.0290  \\
          &       & Suffix & 0.1151  & 0.0648  & 0.1680  & 0.0782  & 0.0283  & 0.0153  & 0.0413  & 0.0186  & 0.0995  & 0.0553  & 0.1485  & 0.0677  & 0.0260  & 0.0143  & 0.0386  & 0.0175  \\
          &       & Silde & 0.2058  & 0.1074  & 0.3174  & 0.1354  & 0.0455  & 0.0234  & 0.0715  & 0.0299  & 0.1969  & 0.1043  & 0.2977  & 0.1296  & 0.0433  & 0.0224  & 0.0683  & 0.0287  \\
          &       & Original & 0.0993  & 0.0553  & 0.1450  & 0.0669  & 0.0054  & 0.0026  & 0.0088  & 0.0034  & 0.1000  & 0.0557  & 0.1430  & 0.0664  & 0.0052  & 0.0026  & 0.0090  & 0.0036  \\
\cmidrule{2-19}          & \multirow{4}[2]{*}{CE} & Prefix & \cellcolor{myred} \textbf{0.2859} & \cellcolor{myred} \textbf{0.1611} & \cellcolor{myred} \textbf{0.3970} & \cellcolor{myred} \textbf{0.1890} & \cellcolor{myred} \textbf{0.0968} & \cellcolor{myred} \textbf{0.0535} & \cellcolor{myred} \textbf{0.1379} & \cellcolor{myred} \textbf{0.0639} & \cellcolor{myblue} \underline{0.2639} & \cellcolor{myblue} \underline{0.1486} & \cellcolor{myred} \textbf{0.3684} & \cellcolor{myblue} \underline{0.1749} & \cellcolor{myred} \textbf{0.0972} & \cellcolor{myred} \textbf{0.0537} & \cellcolor{myred} \textbf{0.1377} & \cellcolor{myred} \textbf{0.0639} \\
          &       & Suffix & 0.1076  & 0.0623  & 0.1455  & 0.0718  & 0.0276  & 0.0151  & 0.0408  & 0.0184  & 0.0970  & 0.0559  & 0.1391  & 0.0664  & 0.0294  & 0.0161  & 0.0425  & 0.0194  \\
          &       & Silde & \cellcolor{myblue} \underline{0.2742} & \cellcolor{myblue} \underline{0.1496} & \cellcolor{myblue} \underline{0.3834} & \cellcolor{myblue} \underline{0.1771} & 0.0806  & 0.0449  & 0.1152  & 0.0537  & \cellcolor{myred} \textbf{0.2704} & \cellcolor{myred} \textbf{0.1512} & \cellcolor{myblue} \underline{0.3679} & \cellcolor{myred} \textbf{0.1757} & \cellcolor{myblue} \underline{0.0853} & \cellcolor{myblue} \underline{0.0475} & \cellcolor{myblue} \underline{0.1216} & \cellcolor{myblue} \underline{0.0566} \\
          &       & Original & 0.1043  & 0.0613  & 0.1427  & 0.0709  & 0.0324  & 0.0183  & 0.0463  & 0.0218  & 0.0977  & 0.0573  & 0.1425  & 0.0686  & 0.0353  & 0.0202  & 0.0503  & 0.0240  \\
    \midrule
    \multirow{8}[4]{*}{Multi} & \multirow{4}[2]{*}{BCE} & Prefix & 0.2440  & 0.1324  & 0.3530  & 0.1598  & 0.0536  & 0.0276  & 0.0830  & 0.0350  & 0.2142  & 0.1123  & 0.3222  & 0.1395  & 0.0586  & 0.0307  & 0.0905  & 0.0388  \\
          &       & Suffix & 0.1719  & 0.0888  & 0.2765  & 0.1152  & 0.0526  & 0.0276  & 0.0800  & 0.0345  & 0.1896  & 0.0999  & 0.2874  & 0.1245  & 0.0605  & 0.0324  & 0.0897  & 0.0398  \\
          &       & Silde & 0.2228  & 0.1196  & 0.3235  & 0.1450  & 0.0419  & 0.0216  & 0.0650  & 0.0274  & 0.2159  & 0.1140  & 0.3293  & 0.1425  & 0.0467  & 0.0238  & 0.0743  & 0.0308  \\
          &       & Original & 0.2108  & 0.1121  & 0.3214  & 0.1399  & 0.0497  & 0.0258  & 0.0772  & 0.0327  & 0.1987  & 0.1041  & 0.2975  & 0.1289  & 0.0543  & 0.0287  & 0.0825  & 0.0358  \\
\cmidrule{2-19}          & \multirow{4}[2]{*}{CE} & Prefix & 0.2659  & 0.1462  & 0.3714  & 0.1727  & 0.0637  & 0.0342  & 0.0939  & 0.0418  & 0.2020  & 0.1077  & 0.2962  & 0.1315  & 0.0631  & 0.0352  & 0.0892  & 0.0418  \\
          &       & Suffix & 0.1964  & 0.1038  & 0.2851  & 0.1260  & 0.0461  & 0.0245  & 0.0681  & 0.0300  & 0.1949  & 0.1053  & 0.2757  & 0.1255  & 0.0524  & 0.0290  & 0.0763  & 0.0350  \\
          &       & Silde & 0.2480  & 0.1376  & 0.3488  & 0.1630  & 0.0733  & 0.0415  & 0.1030  & 0.0490  & 0.2348  & 0.1317  & 0.3325  & 0.1563  & 0.0753  & 0.0425  & 0.1074  & 0.0505  \\
          &       & Original & 0.2654  & 0.1443  & 0.3705  & 0.1709  & \cellcolor{myblue} \underline{0.0901} & \cellcolor{myblue} \underline{0.0507} & \cellcolor{myblue} \underline{0.1276} & \cellcolor{myblue} \underline{0.0602} & 0.2079  & 0.1166  & 0.3050  & 0.1410  & 0.0778  & 0.0438  & 0.1087  & 0.0515  \\
    \bottomrule
     \end{tabular}}}%
    \vspace{-1em}
  \label{tab:when_effective_2}%
\end{table*}%

\subsection{Additional Experimental setup}

\noindent \textbf{Details of the Selected Models.} We select four of the most representative and influential SR models from Table \ref{tab:baseline_results} for exploration in this section. Two are from SR models without SSS, two are from SR models with SSS. These four models, spanning 2016 to 2024 and based on different architectures, rank among the most highly cited models published during that period.
\begin{itemize}
    \item \textbf{GRU4Rec} \cite{hidasi2015session}: This is the first model to apply the Gated Recurrent Unit (GRU) to model sequences of user behavior for SR.
    \item \textbf{SASRec} \cite{kang2018self}: It is a representative Transformer-based model, which leverages the multi-head self-attention for the SR task.
    \item \textbf{FMLPRec} \cite{zhou2022filter}: It is an all-MLP model using a learnable filter-enhanced block to remove noise in the embedding matrix.
    \item \textbf{BSARec} \cite{shin2024attentive}: It is a hybrid model that combines self-attention with the Fourier Transform to address the oversmoothing.
\end{itemize}

\noindent \textbf{Loss Functions and Additional Datasets.} For the loss function, we selected the binary cross-entropy loss (BCE) \cite{hidasi2015session,kang2018self} and cross-entropy loss (CE) \cite{shin2024attentive,shin2025tv}, which are the most commonly used in SR. Given the relatively small scale of the dataset used in Section \ref{sec:effects}, we add two new, larger-scale datasets, each containing nearly 1 million user-item interactions. MovieLens-1M (\textbf{ML-1M}) is a movie recommendation dataset from MovieLens\footnote{https://grouplens.org/datasets/movielens/}. It contains 6,041 users, 3,417 items, and 999,611 interactions. The average sequence length is 165.5, with a sparsity of 95.16\%. The \textbf{CDs} is from the Amazon shopping platform \cite{mcauley2015inferring}, corresponding to the ``CDs and Vinyl'' category. It contains 75,258 users, 66,443 items, and 1,097,592 interactions. The average sequence length is 14.6, with a sparsity of 99.97\%.

\begin{figure}[!t]
  \centering
  \includegraphics[scale=0.325]{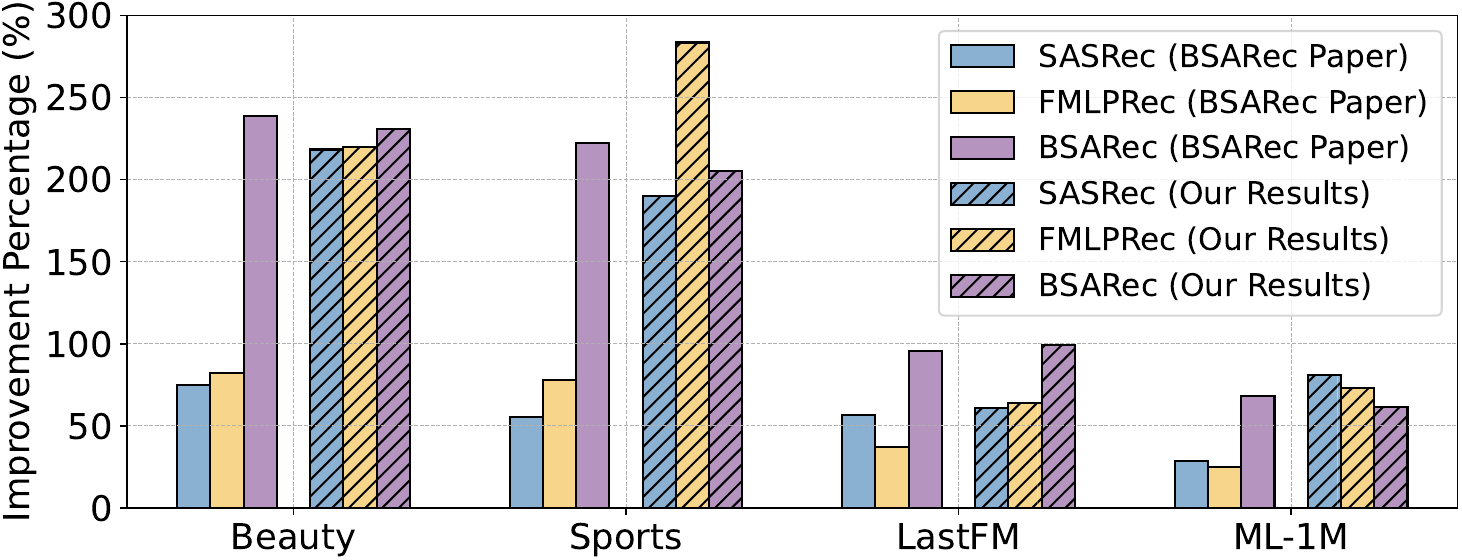}
  \vspace{-1em}
  \caption{Average performance improvements compared to the GRU4Rec reported in the BSARec \cite{shin2024attentive} paper.}
  \vspace{-1em}
  \label{tab:best_in_paper}
\end{figure}

\subsection{Results Analysis}

The performance of four backbone SR models with different loss functions, splitting methods, and target strategies are presented in Table \ref{tab:when_effective_1} (four small-scale datasets) and \ref{tab:when_effective_2} (two large-scale datasets). From the results, we have the following observations and analysis:

(1) Overall, the multi-target strategy outperforms the single-target strategy because the former provides more supervision signals from earlier positions in the sequence, enabling the model to learn user preference knowledge and behavioral patterns more comprehensively. Regarding loss functions, CE loss outperforms BCE loss overall. The former, when paired with the probability distribution of the entire item set output by softmax, supports ranking and better aligns with the multi-class next-item prediction task in sequential recommendation. Among splitting methods, prefix splitting performs better overall than suffix splitting and the sliding window. When employing the single-target strategy (particularly in combination with BCE loss), all three methods achieve significant performance gains over the original model without SSS. Nevertheless, when employing the multi-target strategy with CE loss, all three approaches yield negligible gains and may even degrade performance. This occurs because every interaction potentially eligible as a target participates in training, with ranking optimization applied across the entire item set. Consequently, incorporating SSS offers limited benefits in such scenarios.

(2) From a model perspective, GRU4Rec achieves the best performance on CDs, SASRec wins on ML-1M, FMLPRec achieves the best performance on Sports and LastFM datasets, while BSARec performs best on Beauty and Douyin. Appropriate settings can enable earlier models to outperform the recently proposed models. Furthermore, if we consider only optimal performance, the performance differences are far more minor than those reported in the paper. Taking BSARec as an example \cite{shin2024attentive}, we present the average percentage improvement in model performance reported in BSARec, along with our newly achieved best performance relative to the GRU4Rec model described in the BSARec paper. The results are shown in Table \ref{tab:best_in_paper}\footnote{Following the experimental setup in the BSARec paper \cite{shin2024attentive}, we compared the performance differences of the models across the Beauty, Sports, LastFM, and ML-1M datasets. The BSARec paper did not utilize Douyin and CDs.}. Our results show that the performance gap between models has narrowed significantly. Notably, BSARec's advantage is no longer as pronounced as reported in the paper, further corroborating the interference of SSS in performance evaluation and its immense potential to enhance recommendation accuracy.

(3) The performance variations resulting from different settings applied to the same model are significant. Table \ref{tab:best_and_worst} presents the maximum performance differences (the average improvement of the best case relative to the worst) for each model on each dataset, revealing substantial disparities. On the largest CD dataset, the best performance of the same model outperformed its worst performance by nearly 17 times, with the greatest improvement reaching up to 24 times. Therefore, selecting appropriate target strategies, splitting methods (and loss functions), or conducting comprehensive experiments and evaluations under different settings are necessary.

(4) Finally, we further tally the number of times each setting secured the championship and runner-up positions in Figure \ref{fig:pie_chart}. We observe that single-target, CE loss, with the prefix SSS \& sliding window, achieves the best performance, followed by the two multi-target settings. The remaining settings yield fewer total times. Therefore, in practice, we recommend that researchers prioritize training models using these well-performed settings.

\section{Why is SSS Effective}\label{sec:why}
In this section, we analyze why SSS is effective from the perspective of reshaping the distribution of the training data. 

\begin{figure}[!t]
  \centering
  \includegraphics[scale=0.475]{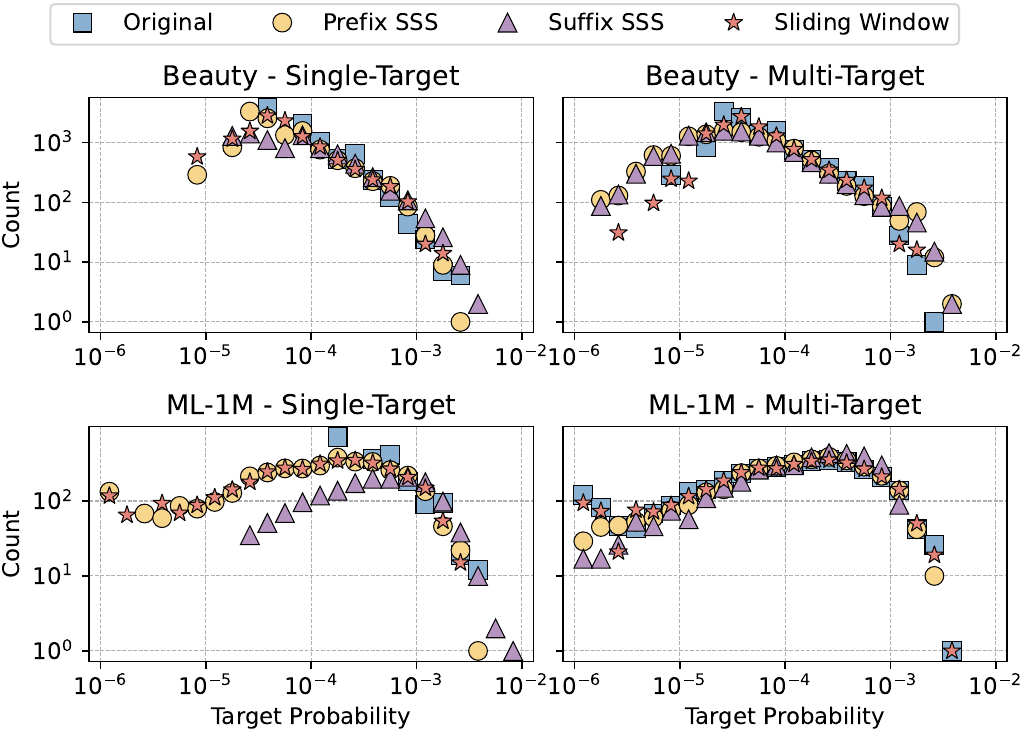}
  \vspace{-1em}
  \caption{Illustration of target distributions with different splitting methods and target strategies.}
  \vspace{-1em}
  \label{fig:target_prob}
\end{figure}

\noindent \textbf{Target Probability Distributions.} Inspired by the previous work \cite{zhou2024contrastive,lee2025sequential}, we visualize the target distributions with different splitting methods and target strategies in Figure \ref{fig:target_prob}, i.e., normalized frequency with which each item appears as the prediction target during training. From the perspective of splitting methods, all three approaches can achieve a more balanced distribution of targets. Without splitting, a few high-frequency items have extremely high probabilities, while most items have extremely low or even zero probabilities. After applying SSS, more items have the opportunity to be selected as targets (the number of circles, triangles, and stars is greater than that of squares). This directly provides the model training with more supervisory signals, enabling it to learn more comprehensive user preferences and transition patterns. From the perspective of target strategies, a multi-target strategy largely compensates for the scarcity of target items that arises from a single-target strategy. Consequently, under the multi-target strategy, the differences of overall target distributions for the original data and the three splitting methods become smaller. However, splitting methods can still improve the target probabilities of different items to some extent.

\begin{table}[!t]
  \centering
  \caption{The maximum performance differences (the average improvement of the best case relative to the worst).}
   \vspace{-1em}
   \setlength{\tabcolsep}{1.1mm}{
 \scalebox{0.9}{
    \begin{tabular}{c|cccccc}
    \toprule
    Mtehod & Beauty & Sports & Douyin & LastFM & ML-1M & CDs \\
    \midrule
    GRU4Rec & 424.53\% & 275.43\% & 460.78\% & 425.16\% & 441.59\% & 2493.27\% \\
    SASRec & 109.63\% & 426.77\% & 74.09\% & 477.22\% & 244.83\% & 1764.68\% \\
    FMLPRec & 118.84\% & 568.24\% & 83.68\% & 595.48\% & 183.88\% & 1724.19\% \\
    BSARec & 106.53\% & 373.54\% & 84.71\% & 697.43\% & 169.59\% & 1709.90\% \\
    \bottomrule
    \end{tabular}}}%
   \vspace{-1em}
  \label{tab:best_and_worst}%
\end{table}%

\vspace{0.3em}

\begin{figure}[!t]
  \centering
  \includegraphics[scale=0.475]{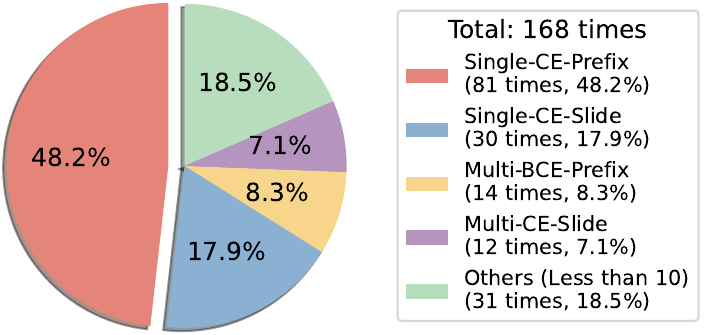}
  \vspace{-1em}
  \caption{The number of times different settings achieve the best performance and the second-best performance.}
  \vspace{-1em}
  \label{fig:pie_chart}
\end{figure}

\begin{figure}[!t]
  \centering
  \includegraphics[scale=0.475]{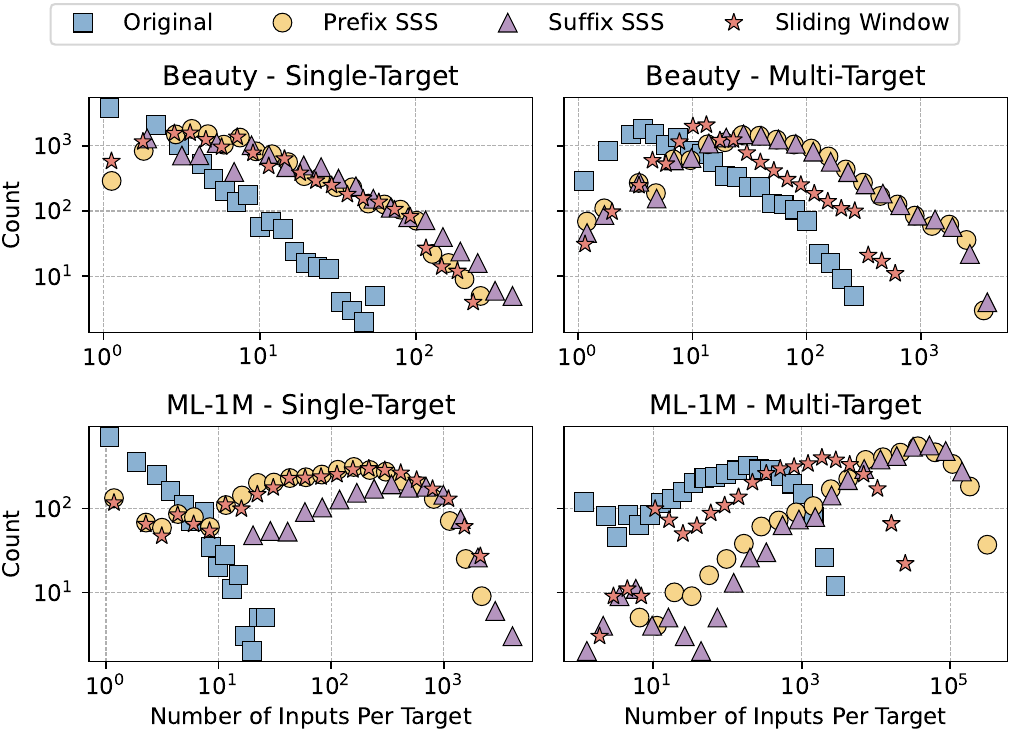}
  \vspace{-1em}
  \caption{Illustration of input-target distributions with different splitting methods and target strategies.}
  \vspace{-1em}
  \label{fig:per_target}
\end{figure}

\noindent \textbf{Input-Target distributions.} Furthermore, we demonstrate the number of inputs associated with each target under different splitting methods and target strategies \cite{lee2025sequential}. The results are presented in Figure \ref{fig:per_target}. We can observe that all three splitting methods significantly increase the number of inputs for each target compared to the original data. This quantitative improvement further enhances the model's ability to generalize the mapping from inputs to targets, enabling it to deliver more accurate recommendation results. Without SSS, numerous items possess only sparse inputs, while only a handful of high-frequency items accumulate substantial input sequences. Consequently, the model can only learn to predict these few high-frequency items, while its prediction performance for the much larger number of low-frequency items remains problematic. Compared with the single-target strategy, the multi-target strategy can further increase the number of inputs per target. This enhances input-target diversity, enabling the model to learn more comprehensive patterns of user preferences. However, as shown in Figure \ref{fig:pie_chart}, the multi-target strategy sometimes performs worse than the single-target strategy. This may be because the former fails to enhance the diversity of input-target pairs beyond the SSS, merely increasing the number of similar pairs. In addition, under the multi-target strategy, both prefix SSS and suffix SSS yield more input-target pairs than the sliding window. However, based on results from Table \ref{tab:when_effective_1} and Table \ref{tab:when_effective_2}, prefix SSS demonstrates significantly superior overall performance compared to suffix SSS. We attribute this to prefix SSS's greater alignment with the temporal accumulation and repetition pattern of user behavior from past to present \cite{ye2020time,li2023repetition}. As this falls outside the work's primary focus, we will continue to explore it in the future.

\section{Discussion}\label{sec:discussion}

\noindent \textbf{The Impact of SSS and Suggestions for Mitigation.} Our work demonstrates how SSS can distort researchers' evaluations of the sequential recommendation models' actual performance. Many recently proposed models may unknowingly incorporate SSS during implementation, leading to falsely high performance. Although our evaluation selected only basic SR backbone models, the impact of SSS may extend far beyond this. Since many SR subdomain methods are based on these fundamental backbones, their implementations may also directly inherit code containing SSS. For example, multi-modal recommendation methods may utilize these models to encode ID sequences \cite{ji2023online,pan2022multimodal}, while cross-domain recommendation models may employ them to capture user preferences within a single domain \cite{lin2024mixed,xu2024rethinking}. The impact of SSS will correspondingly affect our evaluation of the actual performance of them. In light of this situation, we provide the following practical suggestions:
\begin{itemize}
    \item Researchers are advised to carefully examine existing open-source codes and projects they used when implementing their new models, ensuring that all baselines and proposed methods use identical splitting methods, target strategies, and other training settings. These settings should be incorporated into the code with full knowledge of their implications.
    \item Whether SSS is employed, the specific SSS used, and the target strategies (along with other settings) should all be presented as part of the implementation details in the paper. This enables readers to more fairly and rigorously reproduce the results in the paper or use them as a baseline for comparison.
    \item Another option worth considering is conducting a comprehensive evaluation of proposed methods across diverse configurations, particularly for foundational SR backbones. They are more likely to serve as the core building blocks for subsequent work. As shown in Tables \ref{tab:when_effective_1}, \ref{tab:when_effective_2}, and Figure \ref{fig:pie_chart}, the models' performance may vary significantly under different training settings.
\end{itemize}
Following the above suggestions will help reduce the interference of SSS in evaluating the actual performance of models, further preventing its propagation into more ongoing or future research and code projects. Additionally, exploring SSS-related training settings may also help us further unlock the potential of existing models.

\vspace{0.3em}

\noindent \textbf{Insights for Data Augmentation Methods.} The value of data augmentation lies not in merely expanding sample size or applying heuristic perturbations to individual samples, but in strategically reshaping the statistical distribution of training data. This insight has been validated by the recommendation performance and visualization results delivered by SSS. The reason SSS significantly improves model performance lies in its ability to achieve a more balanced target distribution through prefix SSS, suffix SSS, or sliding window. These approaches increases exposure opportunities for low-frequency items while enriching the diversity of input-target pairs (matching more effective contexts to the same target), rather than merely increasing sample quantity. Future data augmentation methods should also focus on optimizing the distribution of training data. This involves learning more comprehensive user preferences by adjusting the bias distribution of original data \cite{chen2021autodebias,yang2023generic}, or adapting to new distributions in the testing phase \cite{wang2022causal,zhao2025distributionally}, rather than limiting efforts to perturbations and augmentations of individual samples. Developing more efficient DA methods based on the analysis in this work is also one of our areas of interest.

\vspace{0.3em}

\noindent \textbf{Beyond SR, Limitations, and Future Work.} In the field of recommender systems, numerous studies have focused on the reproducibility of papers and models. Such work primarily examines whether the results reported in papers can be replicated or whether novel approaches demonstrate progress over classical methods \cite{ferrari2019we,shehzad2025worrying}. However, factors that appear commonplace yet significantly impact model performance (such as SSS in our work) remain largely unexplored. In our work, we focus solely on the impact of SSS, a ubiquitous yet often overlooked factor, on the performance of sequential recommendation models. We also observe that a single factor can lead to phantom progress \cite{ferrari2019we} across numerous models. Therefore, future research should place greater emphasis on investigating these seemingly ordinary factors. These studies should not be confined solely to SR but can be progressively extended to domains such as cross-domain recommendation \cite{zang2022survey}, multi-modal recommendation \cite{liu2024multimodal}, and large language models based recommendation \cite{zhao2024recommender,liu2025large,lan2025efficient,lan2025ept}, etc. Investigating and analyzing these factors will not only facilitate fair and rigorous model evaluation but also potentially aid in discovering and proposing new methodologies.

Additionally, we can observe that some earlier models can outperform the recently proposed models when appropriate training settings are selected. However, our research does not advocate abandoning the study of new models. Previous work has demonstrated that complex models yield significant results in practical applications \cite{anelli2023challenging}, underscoring the need for continued research of new architectures. Moreover, our work does not provide a theoretical analysis of how splitting methods, target strategies, and loss functions influence data distribution and model performance. Consequently, developing a unified theoretical framework that integrates these factors represents another area of interest for us.

\section{Related Work}
\noindent \textbf{Sequential Recommendation.} SR learns user preferences from their historical behavior sequences and provides personalized suggestions \cite{kang2018self,dang2023uniform}. Early works used Markov Chains to predict the next item based on the most recent few interactions \cite{rendle2010factorizing,he2016fusing}. Later, researchers adopted RNNs \cite{2022cmnrec} to capture the relationships among items. GRU4Rec \cite{hidasi2015session} trained a Gated Recurrent Unit (GRU) to model the evolution of user interests. However, RNNs suffer from the drawback of poor parallelizability. Thus, researchers used CNNs to learn sequential patterns \cite{tang2018personalized} or model the dependencies in sequences \cite{yuan2019simple}. More recently, a series of attention-based models emerged due to the success of the Transformer \cite{vaswani2017attention}. SASRec \cite{kang2018self} used multi-head self-attention to learn the importance of different interactions. STOSA \cite{fan2022sequential} embeds each item as a stochastic Gaussian distribution and devised a Wasserstein Self-Attention to characterize item-item relationships. Beyond the Transformer, FMLP-Rec \cite{zhou2022filter} proposed an all-MLP model with learnable filters for SR. BSARec \cite{shin2024attentive} combined self-attention with the Fourier Transform to address the oversmoothing problem. LRURec \cite{yue2024linear} introduced linear recurrence with matrix diagonalization to capture transition patterns.

\vspace{0.3em}

\noindent \textbf{Data Augmentation.} Due to the widespread data sparsity in SR, many DA methods have been proposed and have made impressive progress \cite{dang2025data,dang2026exploring,dang2026tail,zhao2024enhancing}. Heuristic methods generate new samples by randomly perturbing the original sequence. Early works include SSS \cite{tang2018personalized} and Dropout \cite{tan2016improved}, which split a sequence into multiple sub-sequences or discard some items from the original data. With the widespread use of contrastive learning in SR, several DA operators have been proposed. Such as Crop, Mask, Reorder, Insert, and Substitute \cite{liu2021contrastive,xie2022contrastive,liang2025self,sun2023theoretically}. Following this idea, many works improved operators by considering time interval \cite{dang2023ticoserec}, user intents \cite{chen2022intent}, and relevance \& diversity \cite{dang2025augmenting}. Furthermore, ICSRec \cite{qin2024intent} extracted coarse-grain intent signals from all users’ historical interaction (sub-)sequences and then used these signals to construct auxiliary learning objectives for SR. Due to the limited controllability of heuristics, which can yield low-quality augmented data, much works explored generative modules to produce new sequence data \cite{yin2024dataset}. DiffASR \cite{liu2023diffusion} adopted the diffusion model for sequence generation with two guide strategies. ASReP \cite{liu2021augmenting} and BARec \cite{jiang2025improving} employed a reversely transformer to generate pseudo-prior items for short sequences.

Although previous studies have experimentally demonstrated that SSS significantly outperforms other DA methods or analyzed the reasons behind SSS's effectiveness, considering only the target strategy and the sliding window \cite{zhou2024contrastive,lee2025sequential}, the interference of SSS on rigorous model evaluation and its effective conditions have never been revealed or explored. We show that SSS interferes with researchers' evaluation of SR models' actual performance through comprehensive experiments. Furthermore, we analyze the conditions under which SSS is effective and the reasons for its effectiveness. Based on the analysis results, we provide a discussion and insights on rigorous evaluation of SR models and on developing more concise and efficient augmentation methods.

\section{Conclusion}
In this work, we reveal the interference of SSS on the models' actual performance evaluation through extensive experiments and analysis. Furthermore, we explore the conditions under which SSS is effective from the perspectives of splitting methods, target strategies, and loss functions. We found that SSS achieves its maximum efficacy only when specific settings are employed concurrently. Additionally, we analyze why SSS is effective from the perspective of the distribution of training data. It evens out the distribution of training data while increasing the likelihood that different items are targeted. Finally, we provide practical suggestions for addressing SSS interference, discuss the development of new DA methods, the impact  of this work beyond SR, and the limitations of this work.

In addition to the future directions mentioned in Section \ref{sec:discussion}, we also plan to incorporate more types of recommendation methods into our analysis. Over the past decade, the entire community has made progress in code and data sharing, as well as in ensuring reproducibility of results \cite{konstan2013toward,ferrari2019we,shehzad2025worrying}. As an academic community, we still have to strive for greater rigor in research practices.

\begin{acks}
This research is partially supported by the National Natural Science Foundation of China under Grant No. (62576083, 62432003, U25A20431). This research is also supported by the Ministry of Education, Singapore, under its Academic Research Fund (AcRF) Tier 1 grant, and funded through the SMU-SUTD Internal Research Grant Call (SMU-SUTD 2023\_02\_01), and in part by the Ministry of Education, Singapore, under its Academic Research Fund Tier 2 (Award No. MOE-T2EP201230015). The authors greatly appreciate the anonymous reviewers for their valuable comments.
\end{acks}

% \begin{acks}
% \end{acks}

\bibliographystyle{ACM-Reference-Format}
\balance
\bibliography{references}
% \appendix

\end{document}